\documentclass[aps,prd,preprint,nofootinbib]{revtex4}
\usepackage{amsmath,amssymb,,mathbbol,graphics,epsfig,subfigure}
\newcommand{\newc}{\newcommand}
\newc{\bsym}{\boldsymbol}
\newc{\mrm}{\mathrm}
\newc{\ovl}{\overline}
\newc{\ovla}{\overleftarrow}
\newc{\ovra}{\overrightarrow}
\newc{\wtil}{\widetilde}
\newc{\eps}{\epsilon}
\newc{\tri}{\triangle}
\newc{\hc}{\dagger}
\newc{\PD}{\partial}
\newc{\ra}{\rightarrow}
\newc{\rg}{\sqrt{G}}
\newc{\CL}{{\mathcal{L}}}
\newc{\SL}{\!\!\!/}
\newc{\bc}{\begin{center}}
\newc{\ec}{\end{center}}
\newc{\bi}{\begin{itemize}}
\newc{\ei}{\end{itemize}}
\newc{\TR}{\mbox{\sl{Tr}}}

\begin{document}
\title{Testing Realistic Quark Mass Matrices in the Custodial Randall-Sundrum
Model with Flavor Changing Top Decays.}
\author{We-Fu Chang}
\email{wfchang@phys.nthu.edu.tw}
\affiliation{Department of Physics, National Tsing Hua University,
Hsin Chu 300, Taiwan}
\author{John N. Ng}
\email{misery@triumf.ca}
\author{Jackson M. S. Wu}
\email{jwu@triumf.ca}
\affiliation{Theory group, TRIUMF, 4004 Wesbrook Mall, Vancouver, B.C., Canada}

\date{\today}

\begin{abstract}
We study quark mass matrices in the Randall-Sundrum (RS) model with bulk
symmetry $SU(2)_L \times SU(2)_R \times U(1)_{B-L}$. The Yukawa couplings are
assumed to be within an order of magnitude of each other, and perturbative. We
find that quark mass matrices of the symmetrical form proposed by Koide
\textit{et.~al.}~[Y.~Koide, H.~Nishiura, K.~Matsuda, T.~Kikuchi and 
T.~Fukuyama, Phys. Rev. D {\bf 66}, 093006 (2002)] can be accommodated in the 
RS framework with the assumption of hierarchyless Yukawa couplings, but not 
the hermitian Fritzsch-type mass matrices. General asymmetrical mass matrices 
are also found which fit well simultaneously with the quark masses and the
Cabibbo-Kobayashi-Maskawa matrix. Both left-handed (LH) and right-handed (RH)
quark rotation matrices are obtained that allow analysis of flavour changing
decay of both LH and RH top quarks. At a warped down scale of 1.65~TeV, the
total branching ratio of $t \ra Z$ + jets can be as high as
$\sim 5 \times 10^{-6}$ for symmetrical mass matrices and
$\sim 2 \times 10^{-5}$ for asymmetrical ones. This level of signal is within
reach of the LHC.
\end{abstract}

\pacs{11.30.Hv,12.15.Ff,13.85.-t,14.65.Ha}

\maketitle

\section{Introduction}
The idea of extra dimensions is by now a well-known one. It has led to new 
solutions to the gauge hierarchy problem without imposing 
supersymmetry~\cite{ADD98,RS1}, and it has opened up new avenues to attack the
flavour puzzle in the Standard Model (SM). One such application is the seminal
proposal of split fermions by Arkani-Hamed and Schmaltz~\cite{AS} that fermion
mass hierarchy can be generated from the wave function overlap of fermions
located differently in the extra dimension. The split fermion scenario had 
been implemented in both flat extra dimension models~\cite{AS,FSF}, and warped
extra dimension Randall-Sundrum (RS) models~\cite{GN,RSF}. Subsequently, 
phenomenologically successful mass matrices were found in the case of one flat 
extra dimension without much fine tuning of the Yukawa couplings~\cite{CHN}, 
and in the case of warped extra dimensions, realistic fermion masses and 
mixing pattern can be reproduced with almost universal bulk Yukawa 
couplings~\cite{H03,CKY06,MSM06}.

To date, many attempts in understanding the fermion flavour structure
come in terms of symmetries. Fermion mass matrix ansatz with a high degree of
symmetry were constructed to fit simultaneously the observed mass hierarchy
and flavour mixing patterns. It is an interesting question whether in the pure
geometrical setting of the RS framework where there are no flavour symmetries
a priori, such symmetrical forms can arise and arise naturally without fine
tuning of the Yukawa couplings, i.e. whether symmetries in the fermion mass
matrices can be compatible with a natural, hierarchyless Yukawa structure in
the RS framework, and to what degree.

Another interesting and related question is whether or not one can 
experimentally discern if the fermion mass matrices are symmetric in the RS
framework. In the SM, only the left-handed (LH) fermion mixings such as the
Cabibbo-Kobayashi-Maskawa (CKM) mixing matrix is measurable, but not the
right-handed (RH) ones. However, in the RS framework the RH fermion mixings
become measurable through the effective couplings of the gauge bosons to the
fermions induced from the Kaluza-Klein (KK) interactions. If the fermion mass
matrices are symmetric, the LH and RH mixing matrices would be the same. Thus
the most direct way of searching for the effects of these RH mixings would
be through the induced RH fermion couplings in flavour changing processes that
are either not present or very much suppressed in the SM.

In this work we study how well the RS setting serves as a framework for 
flavour physics either with or without symmetries in the fermion mass matrices,
and if the two scenarios can be distinguished experimentally. We concentrate 
on the quark (and especially the top) sector, and we study the issues involved
in the RS1 model~\cite{RS1} with an $SU(2)_L \times SU(2)_R \times U(1)_X$ 
bulk symmetry, which we shall refer to as the minimal custodial RS (MCRS) 
model. The $U(1)_X$ is customarily identified with $U(1)_{B-L}$. The enlarged 
electroweak symmetry contains a custodial isospin symmetry which protects the 
SM $\rho$ parameter from receiving excessive corrections, and the model has 
been shown to be a complete one that can pass all electroweak precision tests 
(EWPT) at a scale of $\sim 3$ to 4~TeV~\cite{ADMS03}. 

The organization of the paper is as follows. In Sec.~\ref{Sec:RSFP} we quickly
review the details of the MCRS model to fix our notations. In
Sec.~\ref{Sec:RSMQ} we investigate which type of mass matrix ansatz is
compatible with Yukawa couplings that are perturbative and not fine-tuned by
matching the ansatz form to that in the MCRS model. Relevant matching formulae
and EWPT limits on the controlling parameters are collected into the two
Appendices. We also investigate possible patterns in the mass matrices by
numerically scanning the EWPT allowed parameter space for those that can
reproduce simultaneously the observed quark masses and the CKM mixing matrix.
In Sec.~\ref{Sec:RHcurr} we study the effects of quark mass matrices being
symmetrical or not are having on flavour changing top decays, $t \ra c(u) Z$,
which are expected to have the clearest signal at the LHC. We summarized our
findings in Sec.~\ref{Sec:Conc}.

\section{\label{Sec:RSFP}Review of the MCRS model}
In this section, we briefly review the set-up of the MCRS model. We summarize
relevant results on the KK reduction and the interactions of the bulk gauge
fields and fermions, and establish the notation to be used below.

\subsection{General set-up and gauge symmetry breaking}
The MCRS model is formulated in a 5-dimensional (5D) background geometry based
on a slice of $AdS_5$ space of size $\pi r_c$, where $r_c$ denotes the radius
of the compactified fifth dimension. Two 3-branes are located at the
boundaries of the $AdS_5$ slice, which are also the orbifold fixed points.
They are taken to be $\phi=0$ (UV) and $\phi=\pi$ (IR) respectively. The
metric is given by
\begin{equation}\label{Eq:metric}
ds^2 = G_{AB}\,dx^A dx^B
= e^{-2\sigma(\phi)}\,\eta_{\mu\nu}dx^{\mu}dx^{\nu}-r_c^2 d\phi^2 \,, \qquad
\sigma(\phi) = k r_c |\phi| \,,
\end{equation}
where $\eta_{\mu\nu} = \mathrm{diag}(1,-1,-1,-1)$, $k$ is the $AdS_5$ 
curvature, and $-\pi\leq\phi\leq\pi$.

The model has $SU(2)_L \times SU(2)_R \times U(1)_{X}$ as its bulk gauge
symmetry group. The fermions reside in the bulk, while the SM Higgs, which is
now a bidoublet, is localized on the IR brane to avoid fine tuning. The 5D
action of the model is given by~\cite{ADMS03}
\begin{equation}\label{Eq:S5D}
S=\int\!d^4x\!\int_{0}^{\pi}\!d\phi\,\sqrt{G}\left[
\CL_g +\CL_f + \CL_{UV}\,\delta (\phi) + \CL_{IR}\,\delta (\phi-\pi)
\right] \,,
\end{equation}
where $\CL_g$ and $\CL_f$ are the bulk Lagrangian for the gauge fields and
fermions respectively, and $\CL_{IR}$ contains both the Yukawa and Higgs
interactions.

The gauge field Lagrangian is given by
\begin{equation}
\CL_g= -\frac{1}{4}\left(
W_{AB}W^{AB} + \wtil{W}_{AB}\wtil{W}^{AB}+\wtil{B}_{AB}\wtil{B}^{AB}
\right) \,,
\end{equation}
where $W$, $\wtil{W}$, $\tilde{B}$ are field strength tensors of $SU(2)_L$,
$SU(2)_R$ and $U(1)_{X}$ respectively.
On the IR brane, $SU(2)_L \times SU(2)_R$ is spontaneously broken down to
$SU(2)_V$ when the SM Higgs acquires a vacuum expectation value (VEV). On the
UV brane, first the custodial $SU(2)_R$ is broken down to $U(1)_R$ by orbifold
boundary conditions; this involves assigning orbifold parities under
$S_1/(Z_2 \times Z^{\prime}_2)$ to the $\mu$-components of the gauge fields:
one assigns $(-+)$ for $\wtil{W}^{1,2}_\mu$, and $(++)$ for all other gauge
fields, where the first (second) entry refers to the parity on the UV (IR)
boundary. Then, $U(1)_R \times U(1)_{X}$ is further broken down to $U(1)_Y$
spontaneously (via a VEV), leaving just $SU(2)_L \times U(1)_Y$ as the unbroken
symmetry group.

\subsection{Bulk gauge fields}
Let $A_M(x,\phi)$ be a massless 5D bulk gauge field, $M = 0,1,2,3,5$. Working
in the unitary gauge where $A_5=0$, the KK decomposition of $A_\mu(x,\phi)$ is
given by (see e.g.~\cite{Pom99,RSF})
\begin{equation}\label{Eq:gKKred}
A_\mu(x,\phi)= \frac{1}{\sqrt{r_c\pi}}\sum_n A_\mu^{(n)}(x)\chi_n(\phi) \,,
\end{equation}
where $\chi_n$ are functions of the general form
\begin{equation}\label{Eq:gWF}
\chi_n = \frac{e^\sigma}{N_n}
\big[J_1(z_n e^\sigma) + b_1(m_n)Y_1(z_n e^\sigma)\big] \,, \qquad
z_n = \frac{m_n}{k} \,,
\end{equation}
that solve the eigenvalue equation
\begin{equation}\label{Eq:gKKeq}
\left(\frac{1}{r_c^2}\PD_\phi\,e^{-2\sigma}\PD_\phi-m_n^2\right)\chi_n = 0 \,,
\end{equation}
subject to the orthonormality condition
\begin{equation}
\frac{1}{\pi}\int^{\pi}_{0}\!d\phi\,\chi_n\chi_m = \delta_{mn} \,.
\end{equation}
Depending on the boundary condition imposed on the gauge field, the
coefficient function $b_1(m_n)$ is given by
\begin{align}
(++)\;\;\mathrm{B.C.}:\quad
b_1(m_n) &= -\frac{J_0(z_n e^{\sigma(\pi)})}{Y_0(z_n e^{\sigma(\pi)})}
= -\frac{J_0(z_n)}{Y_0(z_n)} \,, \\
(-+)\;\;\mathrm{B.C.}:\quad
b_1(m_n) &= -\frac{J_0(z_n e^{\sigma(\pi)})}{Y_0(z_n e^{\sigma(\pi)})}
= -\frac{J_1(z_n)}{Y_1(z_n)} \,,
\end{align}
which in turn determine the gauge KK eigenmasses, $m_n$. For fields with the
$(++)$ boundary condition, the lowest mode is a massless state $A_\mu^{(0)}$
with a flat profile
\begin{equation}\label{Eq:gflat}
\chi_0 = 1 \,,
\end{equation}
while no zero-mode exists if it is the $(-+)$ boundary condition. The SM gauge
boson is identified with the zero-mode of the appropriate bulk gauge field
after KK reduction.

\subsection{Bulk fermions}
The free 5D bulk fermion action can be written as (see e.g.~\cite{RSF,GN})
\begin{equation}
S_f = \int\!d^4x\!\int^\pi_{0}\!d\phi\,\sqrt{G}\left\{
E^M_a\left[\frac{i}{2}\bar{\Psi}\gamma^a(\ovra{\PD_M}-\ovla{\PD_M})\Psi\right]
+m\,\mathrm{sgn}(\phi)\bar{\Psi}\Psi\right\} \,,
\end{equation}
where $\gamma^a = (\gamma^\mu,i\gamma^5)$ are the 5D Dirac gamma matrices in flat
space, $G$ is the metric given in Eq.~\eqref{Eq:metric}, $E^A_a$ the inverse
vielbein, and $m = c\,k$ is the bulk Dirac mass parameter. There is no
contribution from the spin connection because the metric is
diagonal~\cite{GN}. The form of the mass term is dictated by the requirement of
$Z_2$ orbifold symmetry~\cite{GN}.
The KK expansion of the fermion field takes the form
\begin{equation}\label{Eq:PsiKK}
\Psi_{L,R}(x,\phi) = \frac{e^{3\sigma/2}}{\sqrt{r_c\pi}}
\sum_{n=0}^\infty\psi^{(n)}_{L,R}(x)f^n_{L,R}(\phi) \,,
\end{equation}
where the subscripts $L$ and $R$ label the chirality of the fields, and
$f^n_{L,R}$ form two distinct sets of complete orthonormal functions, which
are found to satisfy the equations
\begin{equation}
\left[\frac{1}{r_c}\PD_\phi-\left(\frac{1}{2}+c\right)k\right]f^n_R
= m_n\,e^\sigma f^n_L \,, \qquad
\left[-\frac{1}{r_c}\PD_\phi+\left(\frac{1}{2}-c\right)k\right]f^n_L
= m_n\,e^\sigma f^n_R \,,
\end{equation}
with the orthonormality condition given by
\begin{equation}\label{Eq:fortho}
\frac{1}{\pi}\int^\pi_{0}\!d\phi\,f^{n\star}_{L,R}(\phi)f^m_{L,R}(\phi)
= \delta_{mn} \,.
\end{equation}

Of particular interest are the zero-modes which are to be identified
as SM fermions:
\begin{equation}
f^0_{L,R}(\phi,c_{L,R}) =
\sqrt{\frac{k r_c\pi(1 \mp 2c_{L,R})}{e^{k r_c\pi(1 \mp 2c_{L,R})}-1}}
e^{(1/2 \mp c_{L,R})k r_c\phi} \,,
\end{equation}
where the upper (lower) sign applies to the LH (RH) label. Depending on the
$Z_2$ parity of the fermion, one of the chiralities is projected out. It can be
seen that the LH zero mode is localized towards the the UV (IR) brane if
$c_L > 1/2$ ($c_L < 1/2$), while the RH zero mode is localized towards the the
UV (IR) brane when $c_R < -1/2$ ($c_R > -1/2$).

The higher fermion KK modes have the general form
\begin{equation}\label{Eq:fWF}
f^n_{L,R} = \frac{e^{\sigma}}{N_n}B_{\alpha}(z_n e^\sigma) \,, \qquad
B_{\alpha}(z_n e^\sigma) =
J_{\alpha}(z_n e^\sigma) + b_{\alpha}(m_n)Y_{\alpha}(z_n e^\sigma) \,,
\end{equation}
where $\alpha = |c \pm 1/2|$ with the LH (RH) mode takes the upper (lower)
sign. Depending on the type of the boundary condition a fermion field has, the
coefficient function $b_{\alpha}(m_n)$ takes the form~\cite{ADMS03}
\begin{align}
\label{Eq:bapp}
(++)\;\;\mathrm{B.C.}:\quad
b_{\alpha}(m_n) &=
-\frac{J_{\alpha \mp 1}(z_n e^{\sigma(\pi)})}
{Y_{\alpha \mp 1}(z_n e^{\sigma(\pi)})} =
-\frac{J_{\alpha \mp 1}(z_n)}{Y_{\alpha \mp 1}(z_n)} \,, \\
\label{Eq:bamp}
(-+)\;\;\mathrm{B.C.}:\quad
b_{\alpha}(m_n) &=
-\frac{J_{\alpha \mp 1}(z_n e^{\sigma(\pi)})}
{Y_{\alpha \mp 1}(z_n e^{\sigma(\pi)})} =
-\frac{J_{\alpha}(z_n)}{Y_{\alpha}(z_n)} \,,
\end{align}
and normalization factor can be written as~\cite{ADMS03}
\begin{align}
(++)\;\;\mathrm{B.C.}:\quad
N_n^2 &= \frac{e^{2\sigma(\phi)}}{2k r_c\pi}
B^2_{\alpha}(z_n e^{\sigma(\phi)})\Big|^{\phi=\pi}_{\phi=0} \,, \\
(-+)\;\;\mathrm{B.C.}:\quad
N_n^2 &= \frac{1}{2k r_c\pi}\big[
e^{2\sigma(\pi)}B^2_{\alpha}(z_n e^{\sigma(\pi)})
-B^2_{\alpha \mp 1}(z_n)\big] \,.
\end{align}
The upper sign in the order of the Bessel functions above applies to the LH
(RH) mode when $c_L > -1/2$ ($c_R < 1/2$), while the lower sign applies to the
LH (RH) mode when $c_L < -1/2$ ($c_R > 1/2$). The spectrum of fermion KK masses
is found from the coefficient function relations given by Eqs.~\eqref{Eq:bapp}
and~\eqref{Eq:bamp}.

Now there is an additional $SU(2)_R$ gauge symmetry over the SM in the bulk,
and the fermions have to be embedded into its representations. Below we chose
the simplest way of doing this, viz. the LH fermions are embedded as $SU(2)_R$
singlets, while the RH fermions are doublets~\cite{ADMS03}. Note that since the
$SU(2)_R$ is broken on the UV brane by the orbifold boundary condition, one
component of the doublet under it must be even under the $Z_2$ parity, and the
other odd. This forces a doubling of RH doublets where the upper component,
say the up-type quark, of one doublet, and the lower component of the other
doublet, the down-type, are even.

\subsection{Fermion interactions}
In 5D, the interaction between fermions and a bulk gauge boson is given by
\begin{equation}
S_{f\bar{f}A} = g_5\int\!d^4x\,d\phi\,\sqrt{G}E^M_a\bar{\Psi}\gamma^a A_M\Psi
+\mathrm{h.\,c.} \,,
\end{equation}
where $g_5$ is the 5D gauge coupling constant. After KK reduction, couplings
of the KK modes in the 4D effective theory arise from the overlap of the wave
functions in the bulk. In particular, the coupling of the {\it m}th and
{\it n}th fermion KK modes to the {\it q}th gauge KK mode is given by
\begin{equation}
g^{m\,n\,q}_{f\bar{f}A} =
\frac{g_4}{\pi}\int^\pi_{0}\!d\phi\,f^m_{L,R}f^n_{L,R}\chi_q \,, \qquad
g_4 = \frac{g_5}{\sqrt{r_c\pi}} \,,
\end{equation}
where $g_4 \equiv g_{SM}$ is the 4D SM gauge coupling constant. Note that
since the gauge zero-mode has a flat profile (Eq.~\eqref{Eq:gflat}), by the
orthonormality condition of the fermions wave functions, Eq.~\eqref{Eq:fortho},
only fermions of the same KK level couple to the gauge zero-mode, and the 4D
coupling is simply given by $g^{m\,m\,0}_{f\bar{f}A} = g_4$.

With the Higgs field $\Phi$ localized on the IR brane, the Yukawa interactions
are contained entirely in $\CL_{IR}$ of the 5D action~\eqref{Eq:S5D}.
The relevant action on the IR brane is given by
\begin{equation}
S_\mathrm{Yuk} = \int\!d^4x\,d\phi\,\sqrt{G}\,\delta(\phi-\pi)
\frac{\lambda_{5,ij}}{k r_c}\,\bar{\Psi}_i(x,\phi)\Psi_j(x,\phi)\Phi(x)
+\mathrm{h.\,c.} \,,
\end{equation}
where $\lambda_{5,ij}$ are the dimensionless 5D Yukawa coupling, and $i,j$ the
family indices. Rescaling the Higgs field to $H(x)= e^{-k r_c\pi}\,\Phi(x)$ so
that it is canonically normalized, the effective 4D Yukawa interaction obtained
after spontaneous symmetry breaking is given by
\begin{equation}
S_\mathrm{Yuk} = \int\!d^4x\,v_W\frac{\lambda_{5,ij}}{k r_c\pi}
\sum_{m,n}\bar{\psi}_{iL}^{(m)}(x)\psi_{jR}^{(n)}(x)
f^m_L(\pi,c^L_{i})f^n_R(\pi,c^R_{j}) + \mathrm{h.\,c.} \,,
\end{equation}
where $\langle H \rangle = v_W = 174$~GeV is the VEV acquired by the Higgs
field. The zero modes give rise to the SM mass terms, and the resulting mass
matrix reads
\begin{equation}\label{Eq:RSM}
(M^{RS}_f)_{ij} = v_W\frac{\lambda^f_{5,ij}}{k r_c\pi}
f^0_{L}(\pi,c^{L}_{f_i})f^0_{R}(\pi,c^{R}_{f_j})
\equiv v_W\frac{\lambda^f_{5,ij}}{k r_c\pi}F_L(c^{L}_{f_i})F_R(c^{R}_{f_j}) \,,
\quad f = u,\,d \,,
\end{equation}
where the label $f$ denotes up-type or down-type quark species. Note that the
Yukawa couplings are in general complex, and so take the form
$\lambda^f_{5,ij} \equiv \rho^f_{ij}e^{i\phi_{ij}}$, with
$\rho^f_{ij},\,\phi_{ij}$  the magnitude and the phase respectively.

\section{\label{Sec:RSMQ}Structure of the quark mass matrices}
In this section, we investigate the possible quark flavour structure in the RS
framework. One immediate requirement on the candidate structures is that the
experimentally observed quark mass spectrum and mixing pattern are reproduced.
Another would be that the 5D Yukawa couplings are all of the same order, in
accordance with the philosophy of the RS framework that there is no intrinsic
hierarchy. We also required that constraints from EWPT are satisfied.

To arrive at the candidate structures, we follow two strategies. One is to
start with a known SM quark mass matrix ansatz which reproduces the observed
quark mass spectrum and mixing pattern. The ansatz form is then matched onto
the RS mass matrix to see if the above requirements are satisfied. The other
strategy is to generate RS mass matrices at random and then pick out those that
satisfy the requirements above.~\footnote{This has been tried before in
Ref.~\cite{H03}, but it was done for the case with $m_{KK} > 10$~TeV where
there is a little hierarchy.}

To solve the hierarchy problem, we take $k r_c = 11.7$ and the warped down
scale to be $\tilde{k} = k e^{-k r_c\pi} = 1.65$~TeV.
Since new physics first arise at the TeV scale in the RS framework, it is also
where experimental data are matched to the RS model below. We will assume that
the CKM matrix evolves slowly between $\mu = M_Z$ and $\mu = 1$~TeV so that
the PDG values can be adopted, and we will use the running quark mass central 
values at $\mu = 1$~TeV from Ref.~\cite{XZZ07}.

\subsection{\label{Sec:MMA}Structure from mass matrix ansatz}
In trying to understand the pattern of quark flavour mixing, many ansatz for
the SM quark mass matrices have been proposed over the years. There are two
common types of mass matrix ansatz consistent with the current CKM data.
One type is the Hermitian ansatz first proposed by Fritzsch some time
ago~\cite{Fansatz}, which has been recently updated to better accommodate
$|V_{cb}|$~\cite{FX03}. The other type is the symmetric ansatz proposed by
Koide \textit{et. al.}~\cite{KNMKF02}, which was inspired by the nearly 
bimaximal mixing pattern in the lepton sector.~\footnote{In the SM, because of
the freedom in choosing the RH flavour rotation, quark mass matrices can always
be made Hermitian. But this need not be the case in the RS framework as we show
below.} Using these ansatz as templates, we find that only the Koide-type
ansatz admit hierarchy-free 5D Yukawa couplings; this property is demonstrated
below. That Fritzsch-type ansatz generically lead to hierarchical Yukawa
couplings is shown in Appendix~\ref{app:HermM}.

The admissible ansatz we found takes the form
\begin{equation}\label{Eq:MNM}
M_f = P_f^\hc \hat{M}_f P_f^\hc \,, \quad f = u,\,d,
\end{equation}
where $P_f = \mrm{diag}\{e^{i\delta^f_1},\,e^{i\delta^f_2},\,e^{i\delta^f_3}\}$
is a diagonal pure phase matrix, and
\begin{equation}
\hat{M}_f =
\begin{pmatrix}
\xi_f & C_f & C_f \\
C_f   & A_f & B_f \\
C_f   & B_f & A_f
\end{pmatrix} \,,
\end{equation}
with all entries real and $\xi_f$ much less than all other entries. When
$\xi_f = 0$, the ansatz of Ref.~\cite{KNMKF02} is recovered.

The real symmetric matrix $\hat{M}_f$ is diagonalized by the orthogonal matrix
\begin{equation}\label{Eq:MNOQ}
O_f^\mrm{T} \hat{M}_f O_f =
\begin{pmatrix}
\lambda^f_1 & 0           & 0 \\
0           & \lambda^f_2 & 0 \\
0           & 0           & \lambda^f_3
\end{pmatrix} \,, \quad
O_f =
\begin{pmatrix}
c_f                    & 0                  & s_f \\
-\frac{s_f}{\sqrt{2}} & -\frac{1}{\sqrt{2}} & \frac{c_f}{\sqrt{2}} \\
-\frac{s_f}{\sqrt{2}} & \frac{1}{\sqrt{2}}  & \frac{c_f}{\sqrt{2}}
\end{pmatrix} \,,
\end{equation}
where the eigenvalues are given by
\begin{align}
\lambda_1^f &= \frac{1}{2}\left[
A_f+B_f+\xi_f-\sqrt{8C_f^2+(A_f+B_f-\xi_f)^2}\right] \,, \notag \\
\lambda_2^f &= A_f-B_f \,, \notag \\
\lambda_3^f &= \frac{1}{2}\left[
A_f+B_f+\xi_f+\sqrt{8C_f^2+(A_f+B_f-\xi_f)^2}\right] \,,
\end{align}
and the mixing angles are given by
\begin{equation}
c_f = \sqrt{\frac{\lambda^f_3-\xi_f}{\lambda^f_3-\lambda^f_1}} \,, \quad
s_f = \sqrt{\frac{\xi_f-\lambda^f_1}{\lambda^f_3-\lambda^f_1}} \,.
\end{equation}
Note that the components of $\hat{M}_f$ can be expressed as
\begin{align}\label{Eq:ABC2m}
A_f &= \frac{1}{2}(\lambda_3^f-\lambda_2^f+\lambda_1^f-\xi_f) \,, \notag \\
B_f &= \frac{1}{2}(\lambda_3^f+\lambda_2^f+\lambda_1^f-\xi_f) \,, \notag \\
C_f &= \frac{1}{2}\sqrt{(\lambda_3^f-\xi_u)(\xi_u-\lambda_1^f)} \,.
\end{align}

To reproduce the observed mass spectrum $m_1^f < m_2^f < m_3^f$, the
eigenvalues $\lambda_i^f$, $i = 1,\,2,\,3$, are assigned to be the appropriate
quark masses. For the Koide ansatz (the $\xi_f = 0$ case), it was pointed out
in Ref.~\cite{MN04} that different assignments are needed for the up and down
sectors to fit $|V_{ub}|$ better. Since the ansatz, Eq.~\eqref{Eq:MNM}, is
really a perturbed Koide ansatz, we follow the same assignments here:
\begin{align}\label{Eq:m2ABC}
\lambda^u_1 &= -m^u_1 \,, & \lambda^u_2 &= m^u_2 \,, & \lambda^u_3 &= m^u_3 \,,
\notag \\
\lambda^d_1 &= -m^d_1 \,, & \lambda^d_2 &= m^d_3 \,, & \lambda^d_3 &= m^d_2 \,.
\end{align}

Now since $O_d^\mrm{T}\hat{M}_d\,O_d = \mrm{diag}\{-m^d_1,\,m^d_3,\,m^d_2\}$
for the down-type quarks, to put the eigenvalues into hierarchical order, the
diagonalization matrix becomes $O'_d = O_d\,T_{23}$, where
\begin{equation}
T_{23} =
\begin{pmatrix}
1 & 0 & 0 \\
0 & 0 & 1 \\
0 & 1 & 0
\end{pmatrix} \,.
\end{equation}
The quark mixing matrix is then given by
\begin{equation}
V_\mrm{mix} = O_u^\mrm{T}P_u P_d^\hc O'_d =
\begin{pmatrix}
c_u c_d+\kappa s_u s_d & c_u s_d-\kappa s_u c_d & -\sigma s_u \\
-\sigma s_d            & \sigma c_d             & \kappa \\
s_u c_d-\kappa c_u s_d & s_u s_d+\kappa s_u s_d & -\sigma c_u
\end{pmatrix} \,,
\end{equation}
where
\begin{equation}
\kappa = \frac{1}{2}(e^{i\delta_3}+e^{i\delta_2}) \,, \quad
\sigma = \frac{1}{2}(e^{i\delta_3}-e^{i\delta_2}) \,, \qquad
\delta_i = \delta^u_i-\delta^d_i \,, \quad i = 1,\,2,\,3 \,.
\end{equation}
Without loss of generality, $\delta_1$ is taken to be zero.

The matrix $V_\mrm{mix}$ depends on four free parameters, $\delta_{2,3}$ and
$\xi_{u,d}$. A good fit to the CKM matrix is found by demanding the following
set of conditions:
\begin{equation}\label{Eq:CKMcond}
|\kappa| = |V_{cb}| = 0.04160 \,, \quad
|\sigma|s_u = |V_{ub}| = 0.00401 \,, \quad
|\sigma|s_d = |V_{cd}| = 0.22725 \,,
\end{equation}
and $\delta_{CP} = -(\delta_3+\delta_2)/2 = 59^\circ$. These imply
\begin{equation}\label{Eq:CKMfit}
\delta_2 = -2.55893 \,, \quad
\delta_3 = -0.49944 \,, \quad
\xi_u = 1.36226 \times 10^{-3} \,,\quad
\xi_d = 6.50570 \times 10^{-5} \,,
\end{equation}
which in turn lead to a Jarlskog invariant of $J = 3.16415 \times 10^{-5}$ and
\begin{equation}
|V_\mrm{mix}| =
\begin{pmatrix}
0.97380 & 0.22736 & 0.00401 \\
0.22725 & 0.97294 & 0.04160 \\
0.00816 & 0.04099 & 0.99913
\end{pmatrix} \,,
\end{equation}
both of which are in very good agreement with the globally fitted data.

With $\delta_{u,d}$ determined, so are $\hat{M}_{u,d}$ also. From
Eq.~\eqref{Eq:ABC2m} we have
\begin{align}\label{Eq:ABCnum}
A_u &= 77.32226 \,, & B_u &= 76.77526 & C_u &= 0.43733 \,, \notag \\
A_d &= 1.26269 \,, & B_d &= -1.21731 & C_d &= 7.91684 \times 10^{-3} \,.
\end{align}
Parameters of the RS mass matrix~\eqref{Eq:RSM} can now be solved for by
matching the RS mass matrix onto the ansatz~\eqref{Eq:MNM}. Starting with
$M^{RS}_u$, there are a total of 24 parameters to be determined: six fermion
wave function values, $F_L(c_{Q_i})$ and $F_R(c_{U_i})$, nine Yukawa 
magnitudes, $\rho^u_{ij}$, and nine Yukawa phases, $\phi^u_{ij}$, where
$i,j = 1,\,2,\,3$.~\footnote{We will denote using subscripts $Q$, $U$, and
$D$ respectively, for the left-handed quark doublet, and the right-handed up-
and down-type singlets of $SU(2)_L$.}
Matching $M^{RS}_u$ to $M_u$ results in nine conditions for both magnitudes and
phases. Thus all the up-type Yukawa phases are determined by the three phases
$\delta^u_i$, while six magnitudes are left as free independent parameters.
These we chose to be $F_L(c_{Q_3})$ and $F_R(c_{U_3})$, which are constrained
by EWPT, and $\rho^u_{11}$, $\rho^u_{21}$, $\rho^u_{31}$, $\rho^u_{32}$.

Next we match $M^{RS}_d$ to $M_d$. Since $F_L(c_{Q_i})$ have already been
determined, there are only 21 parameters left in $M^{RS}_d$: $F_R(c_{D_i})$,
$\rho^d_{ij}$, and $\phi^d_{ij}$. Again all the down-type Yukawa phases are
determined by the three phases, $\delta^d_i$, leaving three free magnitudes 
which we chose to be $\rho^d_{31}$, $\rho^d_{32}$, and $\rho^d_{33}$. We 
collect all relevant results from the matching processes into 
Appendix~\ref{app:SymmM}.

To see that the ansatz~\eqref{Eq:MNM} does not lead to a hierarchy in the
Yukawa couplings, note from Eq.~\eqref{Eq:ABC2m} we have
\begin{equation}
A_f \sim |B_f| \sim \frac{m_3^f}{2} \,, \quad
C_u\sim\frac{\sqrt{m_3^u\,m_1^u}}{2} \,, \quad
C_d\sim\frac{\sqrt{m_2^d\,m_1^d}}{2} \,.
\end{equation}
Given this and Eq.~\eqref{Eq:CKMfit}, we see from Eqs.~\eqref{Eq:yQu}
and~\eqref{Eq:yQd} that as long as
\begin{equation}\label{Eq:freerho}
\rho^d_{31}\sim\rho^d_{32}\sim\rho^d_{33}\sim
\rho^u_{11}\sim\rho^u_{21}\sim\rho^u_{31}\sim\rho^u_{32}\sim\rho^u_{33} \,,
\end{equation}
all Yukawa couplings would be of the same order in magnitude. It is amusing to
note that if we begin by imposing the condition that the 5D Yukawa couplings
are hierarchy-free instead of first fitting the CKM data, we find
\begin{equation}
\xi_u \sim m_1^u \sim 10^{-3} \,, \quad
\xi_d \sim \sqrt{m_2^d\,m_1^d}\sqrt{\frac{m_1^u}{m_3^u}}
\sim 3 \times 10^{-5} \,,
\end{equation}
which give the correct order of magnitude for $\xi_{u,d}$ necessary for
$V_\mrm{mix}$ to fit the experimental CKM values.

From relations~\eqref{Eq:yQu}, \eqref{Eq:FLQFRU}, and~\eqref{Eq:FRD}, for mass
matrices given by the ansatz~\eqref{Eq:MNM}, all localization parameters can
be determined from just that of the third generation $SU(2)_L$ doublet,
$c_{Q_3}$, and the Yukawa coupling magnitudes listed in Eq.~\eqref{Eq:freerho}.
To satisfy the bounds from flavour-changing neutral-currents (FCNCs), LH light 
quarks from the first two generations should be localized towards the UV brane. 
As discussed in Appendix~\ref{app:SymmM}, for generic choices of Yukawa
couplings this is so for the first generation LH quarks, but not for the second 
generation. In order to have $c_{Q_2} > 0.5$ while still satisfying
constraints from Eqs.~\eqref{Eq:constr2} and~\eqref{Eq:constr3} and the EWPT
constraint $c_{U_3} < 0.2$, we choose
\begin{equation}\label{Eq:UVchoice}
\frac{\rho^u_{31}}{\rho^u_{21}} = 0.2615 \,, \quad
\rho^u_{11} = \rho^u_{31} = 0.7 \,, \quad \rho^u_{33} = 0.85 \,, \quad
\rho^u_{32} = \rho^d_{31} = \rho^d_{31} = \rho^d_{33} = 1 \,.
\end{equation}
We also have to shorten the EWPT allowed range of $c_{Q_3}$ to $(0.3,0.4)$ so
that $c_{Q_2} > 0.5$ is always satisfied. Note that relation~\eqref{Eq:FLQFRU}
constrains $c_{U_2}$ to be greater than $-0.5$ if the perturbativity
constraint, $\lambda_5 < 4$, is to be met.

The localization parameters increase monotonically as $c_{Q_3}$ increases.
Except for $c_{U_{2,3}}$, the variation of the localization parameters is 
small. We list below their range variation as $c_{Q_3}$ varies from 0.3 to 0.4
given the choice of the Yukawa couplings~\eqref{Eq:UVchoice}:
\begin{gather}
0.65 < c_{Q_1} < 0.66 \,, \qquad 0.50 < c_{Q_2} < 0.52 \,, \notag \\
-0.62 < c_{U_1} < -0.61 \,, \qquad -0.26 < c_{U_2} < -0.01 \,, \qquad
-0.16 < c_{U_3} <  0.18 \,, \notag \\
-0.75 < c_{D_1} < -0.74 \,, \qquad -0.60 < c_{D_{2,3}} < -0.59 \,.
\end{gather}

\subsection{\label{Sec:Rand}Structure from numerical search}
The RS mass matrix given by Eq.~\eqref{Eq:RSM} has a productlike form:
\begin{equation}
M^{RS}\sim
\begin{pmatrix}
a_1 b_1 & a_1 b_2 & a_1 b_3 \\
a_2 b_1 & a_2 b_2 & a_2 b_3 \\
a_3 b_1 & a_3 b_2 & a_3 b_3
\end{pmatrix} \,, \qquad
a_i = F_L(c^L_i) \,, \quad b_i = F_R(c^R_i) \,,
\end{equation}
and it can be brought into the diagonal form by a unitary transformation
\begin{equation}
(U_L^f)^\hc M^{RS}_f\,U_R^f =
\begin{pmatrix}
\lambda^f_1 & 0           & 0 \\
0           & \lambda^f_2 & 0 \\
0           & 0           & \lambda^f_3
\end{pmatrix} \,, \quad f = u,\,d \,.
\end{equation}
Suppose there is just one universal 5D Yukawa coupling, say $\lambda_5 = 1$,
then the RS mass matrix $M^{RS}_f$ would be singular with two zero
eigenvalues, and both the LH and RH quark mixing matrices would be the
identity matrix, i.e. 
$V^{L,R}_{mix} = (U^u_{L,R})^\hc U^d_{L,R} = \mathbb{1}_{3 \times 3}$. Thus, in
order to obtain realistic quark masses and CKM mixing angles
($V^L_{mix} \equiv V_{CKM}$), one cannot assume one universal Yukawa coupling.
Rather, for each configuration of localization parameters, the magnitudes and 
phases of the 5D Yukawa coupling constants, $\rho_{ij}$ and $\phi_{ij}$, will 
be randomly chosen from the intervals $[1.0,3.0]$ and $[0,2\pi]$ respectively,
and we take a sample size of $10^5$.

The numerical search is done with $0.5 < c_{Q_{1,2}} < 1$ and
$-1 < c_{U_{1,2}},\,c_{D_{1,2,3}} < -0.5$ so that the first two generation
quarks, as well as the third generation RH quarks of the $D_3$ doublet are
localized towards the UV brane. For the third generation, 
$0.25 < c_{Q_3} < 0.4$ and $-0.5 < c_{U_3}< 0.2$ are required so the EWPT
constraints are satisfied (see Appendix~\ref{app:SymmM}). We averaged the 
quark masses and CKM mixing angles over the entire sample for each 
configuration of localization parameters, and these choices yielded averaged 
values that are within one statistical deviation of the experimental values at
$\mu = 1$~TeV as given in Ref.~\cite{XZZ07}. Below we give three 
representative configurations from the the admissible configurations found 
after an extensive search. 
\begin{itemize}
\item Configuration~I:
\begin{align}
c_Q &= \{0.634,0.556,0.256\} \,, \notag \\
c_U &= \{-0.664,-0.536,0.185\} \,, \notag \\
c_D &= \{-0.641,-0.572,-0.616\} \,.
\end{align}
\end{itemize}
In units of GeV, the mass matrices averaged over the whole sample are given by
\begin{equation}
\langle|M_u|\rangle =
\begin{pmatrix}
8.97\times 10^{-4} & 0.049 & 0.767 \\
0.010              & 0.554 & 8.69 \\
0.166              & 9.06  & 142.19
\end{pmatrix} \,, \quad
\langle|M_d|\rangle =
\begin{pmatrix}
0.0019 & 0.017 & 0.0044 \\
0.022  & 0.196 & 0.050 \\
0.352  & 3.209 & 0.813
\end{pmatrix} \,,
\end{equation}
which have eigenvalues
\begin{align}
m_t &= 109(52) \,, &
m_c &= 0.56(59) \,, &
m_u &= 0.0011(12) \,, \notag \\
m_b &= 2.59 \pm 1.11 \,, &
m_s &= 0.048(32) \,, &
m_d &= 0.0017(12) \,.
\end{align}
The resulting mixing matrices are given by
\begin{align}
|V^{L}_{us}| &= 0.16(14) \,, &
|V^{L}_{ub}| &= 0.009(11) \,, &
|V^{L}_{cb}| &= 0.079(74) \,, \notag \\
|V^{R}_{us}| &= 0.42(24) \,, &
|V^{R}_{ub}| &= 0.12(10) \,, &
|V^{R}_{cb}| &= 0.89(13) \,,
\end{align}
which give rise to an averaged Jarlskog invariant consistent with zero with a
standard error of $1.3 \times 10^{-4}$.

\begin{itemize}
\item Configuration~II:
\begin{align}
c_Q &= \{0.629,0.546,0.285\} \,, \notag \\
c_U &= \{-0.662,-0.550,0.080\} \,, \notag \\
c_D &= \{-0.580,-0.629,-0.627\} \,.
\end{align}
\end{itemize}
In units of GeV, the mass matrices averaged over the entire sample are given by
\begin{equation}
\langle|M_u|\rangle =
\begin{pmatrix}
0.0011 & 0.039 & 0.834 \\
0.014  & 0.492 & 10.55 \\
0.16   & 5.726 & 122.87
\end{pmatrix} \,, \quad
\langle|M_d|\rangle =
\begin{pmatrix}
0.017 & 0.0034 & 0.0036 \\
0.209 & 0.043  & 0.046 \\
2.43  & 0.506  & 0.539
\end{pmatrix} \,,
\end{equation}
which have eigenvalues
\begin{align}
m_t &= 95(45) \,, &
m_c &= 0.49(50) \,, &
m_u &= 0.0014(16) \,, \notag \\
m_b &= 2.01(83) \,, &
m_s &= 0.057(35) \,, &
m_d &= 0.0022(15) \,.
\end{align}
The resulting mixing matrices are given by
\begin{align}
|V^{L}_{us}| &= 0.14(12) \,, &
|V^{L}_{ub}| &= 0.011(13) \,, &
|V^{L}_{cb}| &= 0.11(10) \,, \notag \\
|V^{R}_{us}| &= 0.30(20) \,, &
|V^{R}_{ub}| &= 0.90(12) \,, &
|V^{R}_{cb}| &= 0.23(15) \,,
\end{align}
which give rise to an averaged Jarlskog invariant consistent with zero with a
standard error of $2.3 \times 10^{-4}$.

\begin{itemize}
\item Configuration~III:
\begin{align}
c_Q &= \{0.627,0.571, 0.272\} \,, \notag \\
c_U &= \{-0.518,-0.664,0.180\} \,, \notag \\
c_D &= \{-0.576,-0.610,-0.638\} \,,
\end{align}
\end{itemize}
In units of GeV, the mass matrices averaged over the entire sample are given by
\begin{equation}
\langle|M^u|\rangle =
\begin{pmatrix}
0.092 & 0.0010 & 0.940 \\
0.554 & 0.0065 & 5.66 \\
13.4  & 0.158  & 136.9
\end{pmatrix} \,, \quad
\langle|M^d|\rangle =
\begin{pmatrix}
0.019 & 0.0066 & 0.0026 \\
0.114 & 0.039  & 0.016 \\
2.774 & 0.955  & 0.376
\end{pmatrix} \,,
\end{equation}
which have eigenvalues
\begin{align}
m_t &= 106(50) \,, &
m_c &= 0.56(55) \,, &
m_u &= 0.0013(12) \,, \notag \\
m_b &= 2.32(94) \,, &
m_s &= 0.036(21) \,, &
m_d &= 0.0023(16) \,.
\end{align}
The resulting mixing matrices are given by
\begin{align}
|V^{L}_{us}| &= 0.27(19) \,, &
|V^{L}_{ub}| &= 0.010(10) \,, &
|V^{L}_{cb}| &= 0.048(44) \,, \notag \\
|V^{R}_{us}| &= 0.77(19) \,, &
|V^{R}_{ub}| &= 0.36(21) \,, &
|V^{R}_{cb}| &= 0.85(15) \,,
\end{align}
which give rise to an averaged Jarlskog invariant consistent with zero with a
standard error of $1.9 \times 10^{-4}$.

In summary, from the numerical study we found that in the RS framework, there
is neither a preferred form for the mass matrix nor a universal RH mixing
pattern. Note that the RH mixing matrix is in general quite different from its
LH counterpart, viz. the CKM matrix.

\section{\label{Sec:RHcurr}Flavour violating top quark decays}
In this section we study the consequences the different forms of quark mass
matrices have on FCNC processes. We focus below on the decay,
$t \ra c\,(u)\,Z \ra c\,(u)\,l\bar{l}$, where $l=e,\mu,\tau,\nu$. Modes which
decay into a real $Z$ and $c\,(u)$-jets are expected to have a much higher
rate than those involving a photon or a light Higgs, which happen through loop
effects. Moreover, much cleaner signatures at the LHC can be provided by
leptonic $Z$-decays.

\subsection{\label{Sec:treeFC}Tree-level flavour violations in MCRS}
Tree-level FCNCs are generic in extra-dimensional models, for both a flat
background geometry~\cite{FCNC1} and a warped one~\cite{RSFCNC,H03,APS05}. 
Because of the KK interactions, the couplings of the $Z$ to the fermions are 
shifted from their SM values. These shifts are not universal in general, and so 
flavour violations necessarily result when the fermions are rotated from the 
weak to the mass eigenbasis.

More concretely, consider the $Z f\bar{f}$ coupling in the weak eigenbasis:
\begin{gather}\label{Eq:Zqq}
\mathcal{L}_\mathrm{NC}\supset g_Z Z_\mu\left\{
Q_Z(f_L)\sum_{i,j}(\delta_{ij}+\kappa_{ij}^L)\bar{f}_{iL}\gamma^\mu f_{jL}+
Q_Z(f_R)\sum_{i,j}(\delta_{ij}+\kappa_{ij}^R)\bar{f}_{iR}\gamma^\mu f_{jR}
\right\} \,,
\end{gather}
where $i$, $j$ are family indices, 
$\kappa_{ij}=\mathrm{diag}(\kappa_1,\kappa_2,\kappa_3)$, and
\begin{equation}
Q_Z(f) = T^3_L(f)-s^2 Q_f \,, \qquad
Q_f = T^3_L(f) + T^3_R(f) + Q_{X}(f) = T^3_L(f) + \frac{Y_f}{2} \,,
\end{equation}
with $Q_f$ is the electric charge of the fermion, $Y_f/2$ the hypercharge,
$T_{L,R}(f)$ the weak isospin under $SU(2)_{L,R}$, and $Q_X(f)$ the charge
under $U(1)_X$. We define $\kappa_{ij}\equiv\delta g^{L,R}_{i,j}/g_Z$ to be 
the shift in the weak eigenbasis $Z$ couplings to fermions relative to its SM 
value given by $g_Z\equiv e/(sc)$, as well as the usual quantities
\begin{equation}
e = \frac{g_L\,g'}{\sqrt{g_L^2+g'\,^2}} \,, \qquad
g' = \frac{g_R\,g_{X}}{g_R^2+g_{X}^2} \,, \qquad
s = \frac{e}{g_L} \,, \qquad c = \sqrt{1-s^2} \,,
\end{equation}
where $g_L = g_{5L}/\sqrt{r_c\pi}$ is the 4D gauge coupling constant of
$SU(2)_L$ (and similarly for the rest). Rotating to the mass eigenbasis of the
SM quarks defined by $f' = U^\dag f$, where the unitary matrix $U$
diagonalizes the SM quark mass matrix, flavour off-diagonal terms appear:
\begin{equation}
\mathcal{L}_\mathrm{FCNC}\supset g_Z Z_\mu\left\{
Q_Z(f_L)\sum_{a,b}\hat{\kappa}_{ab}^L\,\bar{f}'_{aL}\gamma^\mu f'_{bL}+
Q_Z(f_R)\sum_{a,b}\hat{\kappa}_{ab}^R\,\bar{f}'_{aR}\gamma^\mu f'_{bR}
\right\} \,,
\end{equation}
where the mass eigenbasis flavour off-diagonal couplings are given by
\begin{equation}\label{Eq:kFCNC}
\hat{\kappa}_{ab}^{L,R} =
\sum_{i,j}(U^\dag_{L,R})_{ai}\kappa_{ij}^{L,R}(U_{L,R})_{jb} \,.
\end{equation}
Note that the off-diagonal terms would vanish only if $\kappa$ is proportional
to the identity matrix.

\begin{figure}[htbp]
\centering
\includegraphics[width=1.5in]{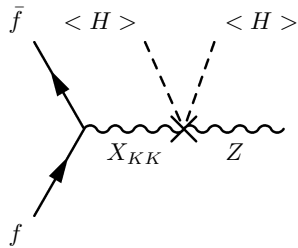}
\caption{\label{Fig:ZKK} Correction to the $Z f\bar{f}$ coupling due to the
exchange of gauge KK modes. The fermions are in the weak eigenbasis, and
$X = Z,\,Z'$.}
\end{figure}

In the RS framework, one leading source of corrections to the SM neutral
current interaction comes from the exchanges of heavy KK neutral gauge bosons
as depicted in Fig.~\ref{Fig:ZKK}.
The effect of gauge KK exchanges give rise only to the diagonal terms of
$\kappa$. It can be efficiently calculated with the help of the massive gauge
5D mixed position-momentum space propagators, which automatically sums up
contributions from all the KK modes~\cite{ADMS03,CDPTW03}.

The leading contributions can be computed in terms of the overlap integral,
\begin{equation}
G_f^{L,R}(c_{L,R}) = \frac{v_W^2}{2}\,r_c\!\int_0^{\pi}\!d\phi
|f^0_{L,R}(\phi,c_{L,R})|^2\tilde{G}_{p=0}(\phi,\pi) \,,
\end{equation}
where $\tilde{G}_{p=0}$ is the zero-mode subtracted gauge propagator evaluated
at zero 4D momentum. For KK modes obeying the $(++)$ boundary condition,
$\tilde{G}_{p=0}$ is given
by~\cite{CDPTW03}
\begin{align}
\tilde{G}^{(++)}_{p=0}(\phi,\phi') = \frac{1}{4k(k r_c\pi)}\bigg\{
\frac{1-e^{2k r_c\pi}}{k r_c\pi}+e^{2k r_c\phi_<}(1-2k r_c\phi_<)
+e^{2k r_c\phi_>}\Big[1+2k r_c(\pi-\phi_>)\Big]\bigg\} \,,
\end{align}
and those obeying the $(-+)$ boundary condition
\begin{equation}
\tilde{G}^{(-+)}_{p=0}(\phi,\phi') =
-\frac{1}{2k}\left(e^{2k r_c\phi_<}-1\right) \,,
\end{equation}
where $\phi_<$ ($\phi_>$) is the minimum (maximum) of $\phi$ and $\phi'$. The
gauge KK correction to the $Z$ coupling is thus
$\kappa^g_{ij} = \kappa^g_{q_i}\delta_{ij}$, with $\kappa^g_{q_i}$ given
by~\cite{CPSW06}
\begin{equation}\label{Eq:dgig}
(\kappa^g_{q_i})_{L,R} = \frac{e^2}{s^2 c^2}\left\{
G^{q^i_{L,R}}_{++}-\frac{G^{q^i_{L,R}}_{-+}}{Q_Z(q^i_{L,R})}\left[
\frac{g_R^2}{g_L^2}c^2 T_R^3(q^i_{L,R})-s^2\frac{Y_{q^i_{L,R}}}{2}\right]
\right\} \,,
\end{equation}
where the label $q$ denotes the fermion species. Note that when the fermions
are localized towards the UV brane ($c_L \gtrsim 0.6$ and $c_R \lesssim -0.6$),
$G_{-+}$ is negligible, while $G_{++}$ becomes essentially flavour
independent~\cite{CPSW06}.

\begin{figure}[htbp]
\centering
\includegraphics[width=2.2in]{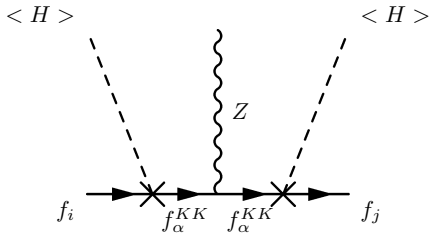}
\caption{\label{Fig:fKK} Correction to the $Z f\bar{f}$ coupling due to SM
fermions mixing with the KK modes. The fermions are in the weak eigenbasis. }
\end{figure}

Another source of corrections to the $Z f\bar{f}$ coupling arises from the
mixings between the fermion zero modes and the fermion KK modes brought about
by the the Yukawa interactions. These generate diagonal as well as off-diagonal
terms in $\kappa$. The diagram involved is depicted in Fig.~\ref{Fig:fKK}.

The effects of the fermion mixings may be similarly calculated by using the
fermion analogue of the gauge propagators. It is however much more convenient
to deal directly with the KK modes here. The KK fermion corrections to the weak
eigenbasis $Z$ couplings can be written as
\begin{equation}\label{Eq:dgfg}
(\kappa^f_{ij})_L = \sum_\alpha\sum_{n=1}^\infty
\frac{m_{i\alpha}^\ast m_{j\alpha}}{(m^\alpha_n)^2}\mathfrak{F}^\alpha_R \,, 
\qquad
(\kappa^f_{ij})_R = \sum_\alpha\sum_{n=1}^\infty
\frac{m_{\alpha i}m_{\alpha j}^\ast}{(m^\alpha_n)^2}\mathfrak{F}^\alpha_L \,,
\end{equation}
where $m_n$ is the $n$th level KK fermion mass, $m_{i\alpha}$ are entries of
the weak eigenbasis RS mass matrix~\eqref{Eq:RSM} with $\alpha$ a generation 
index~\footnote{For shift in the LH couplings, the index $\alpha$ runs over 
the generations of both types of $SU(2)_R$ doublets, $U$ and $D$, both of 
which contain KK modes that can mix with LH zero modes. For shift in
the RH couplings, $\alpha$ runs over just the generations of the only type of 
$SU(2)_L$ doublets, $Q$.}, and 
\begin{equation}
\mathfrak{F}^\alpha_{R,L} = \bigg|\frac{f^n_{R,L}(\pi,c_\alpha^{R,L})}
{f^0_{R,L}(\pi,c_\alpha^{R,L})}\bigg|^2
\frac{Q_Z(f_{R,L})}{Q_Z(f_{L,R})} \,,
\end{equation}
with the argument of $Q_Z$, $f = u,\,d$, denoting up-type or down-type quark
species. Note that for $c_\alpha^L < 1/2$ and
$c_\alpha^R > -1/2$, $|f^n_{L,R}(\pi,c_\alpha^{L,R})|\approx\sqrt{2k r_c\pi}$.

To determine $\hat{\kappa}_{ab}$ in Eq.~\eqref{Eq:kFCNC}, one needs to know
the rotation matrices $U_L$ and $U_R$. In the case where the weak eigenbasis
mass matrices are given by the symmetric ansatz~\eqref{Eq:MNM}, the analytical
form of the rotation matrices are known. By rephasing the quark fields so that
$\delta^u_i = 0$ and all the Yukawa phases reside in down sector, the up-type
rotation matrix is just the orthogonal diagonalization matrix given by
Eq.~\eqref{Eq:MNOQ}. Using the solution of the CKM fit given in
Eq.~\eqref{Eq:CKMfit}, we have
\begin{equation}
U^u_L = U^u_R = U^u \,, \qquad
U^u = O_u =
\begin{pmatrix}
0.99999  & 0                   & 0.00401 \\
-0.00284 & -\frac{1}{\sqrt{2}} & 0.70710 \\
-0.00284 & \frac{1}{\sqrt{2}}  & 0.70710
\end{pmatrix} \,.
\end{equation}

Since we are interested in flavour violating top decays, the relevant mass 
eigenbasis off-diagonal corrections are
$\hat{\kappa}_{3r} = \hat{\kappa}^g_{3r} + \hat{\kappa}^f_{3r}$, $r = 1,\,2$.
For the discussion below, using relations~\eqref{Eq:yQu}, \eqref{Eq:FLQFRU},
and~\eqref{Eq:FRD} we will trade the dependences of $\hat{\kappa}^{L,R}_{ab}$
on all the different localization parameters for just a single dependence on
$c_{Q_3}$, and the Yukawa coupling magnitudes which we fix to take the values
given in Eq.~\eqref{Eq:UVchoice}. Recall that with this choice of the Yukawa
coupling magnitudes, the EWPT allowed range for $c_{Q_3}$ is between 0.3 and
0.4.

\begin{figure}[htbp]
\centering
\subfigure[]{
\label{Fig:subfig:gKKtu}
\includegraphics[width=2.8in]{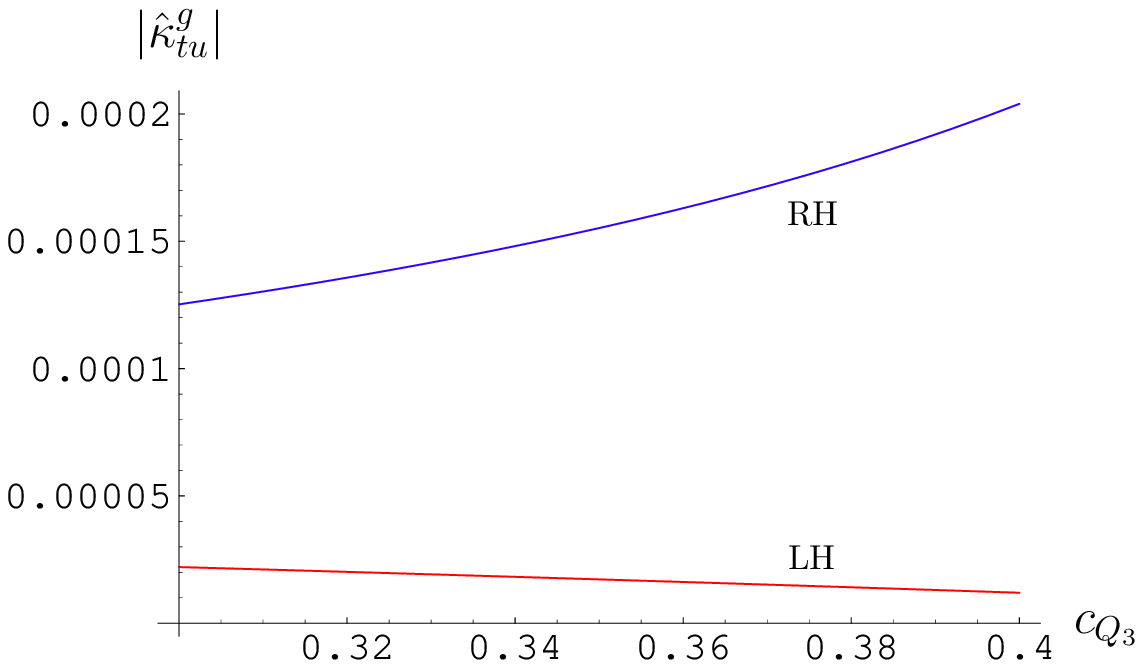}}
\hspace{0.2in}
\subfigure[]{
\label{Fig:subfig:gKKtc}
\includegraphics[width=2.8in]{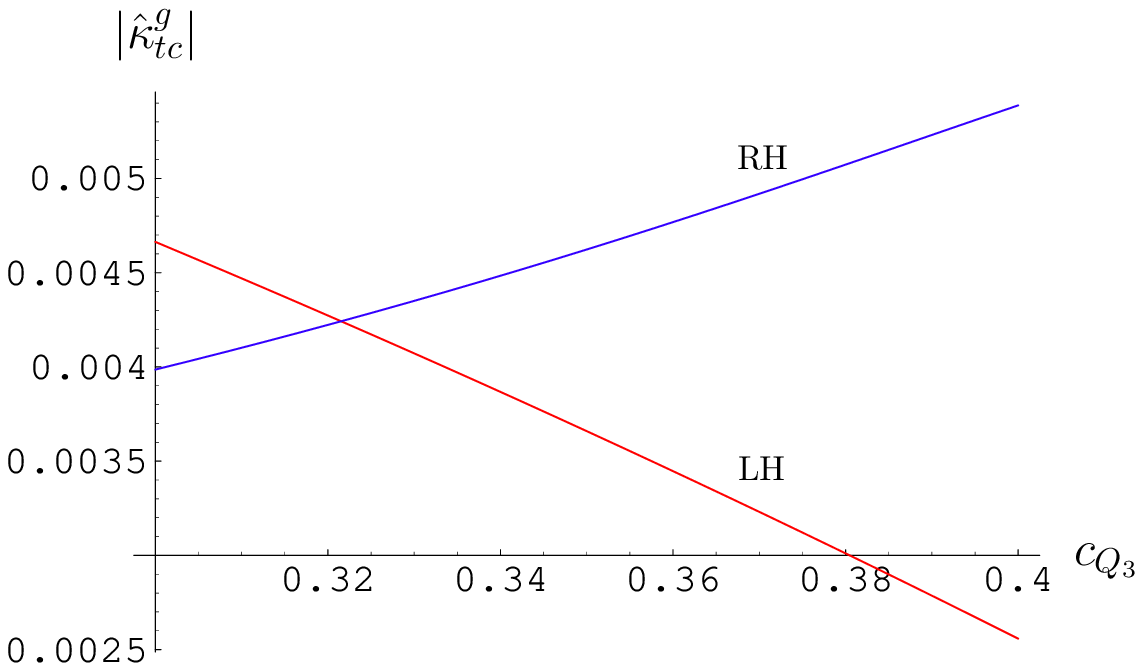}}
\caption{\label{Fig:hkgtuc} Gauge KK contribution in the case of symmetrical 
mass matrices to (a) $\hat{\kappa}_{tu}$ and (b) $\hat{\kappa}_{tc}$. The 
labels LH and RH indicate whether it is for the LH or RH coupling.}
\end{figure}

Since $\kappa^g_{ij} = \kappa^g_{q_i}\delta_{ij}$, the gauge KK contributions
is simply
$\hat{\kappa}^g_{3r} = \sum_i\kappa^g_{q_i}(U^u)^\dag_{3i}U^u_{ir}$, with
\begin{equation}\label{Eq:hatkg}
\hat{\kappa}^g_{tu} =
2.00672 \times 10^{-3}\,(2\kappa^g_{u}-\kappa^g_{c}-\kappa^g_{t}) \,, \qquad
\hat{\kappa}^g_{tc} = 0.50\,(\kappa^g_{t}-\kappa^g_{c}) \,.
\end{equation}
We plot $\hat{\kappa}^g_{3r}$ 
as a function of $c_{Q_3}$ in Fig.~\ref{Fig:hkgtuc}. 

\begin{figure}[htbp]
\centering
\subfigure[]{
\label{Fig:subfig:fKKtu}
\includegraphics[width=2.8in]{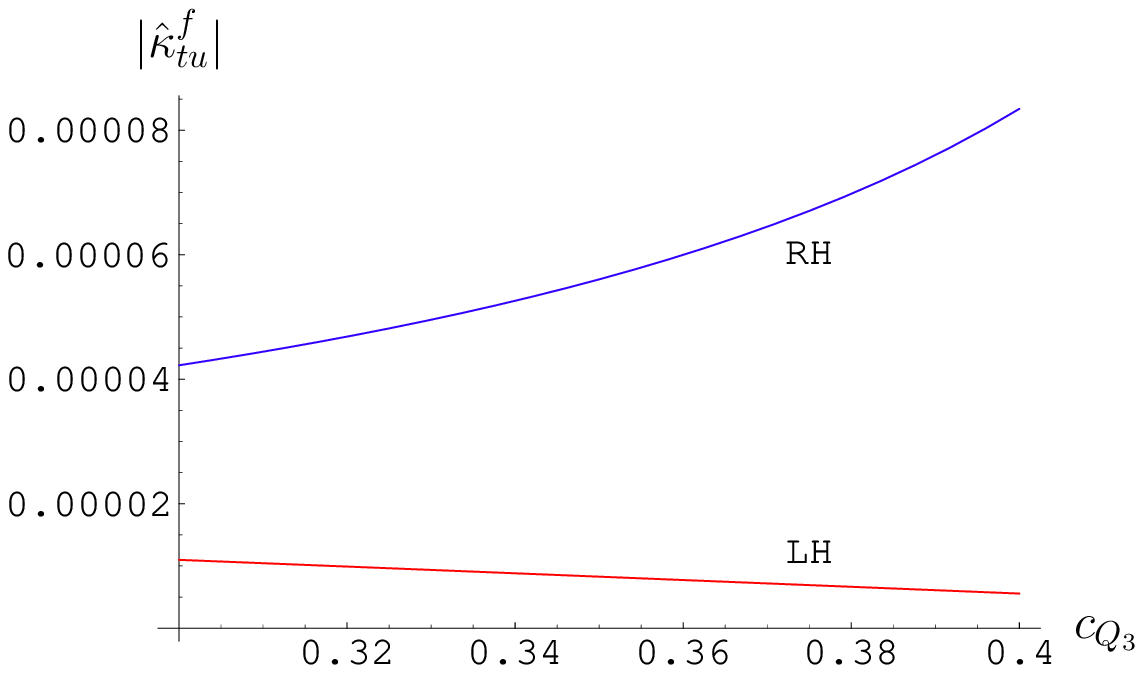}}
\hspace{0.2in}
\subfigure[]{
\label{Fig:subfig:fKKtc}
\includegraphics[width=2.8in]{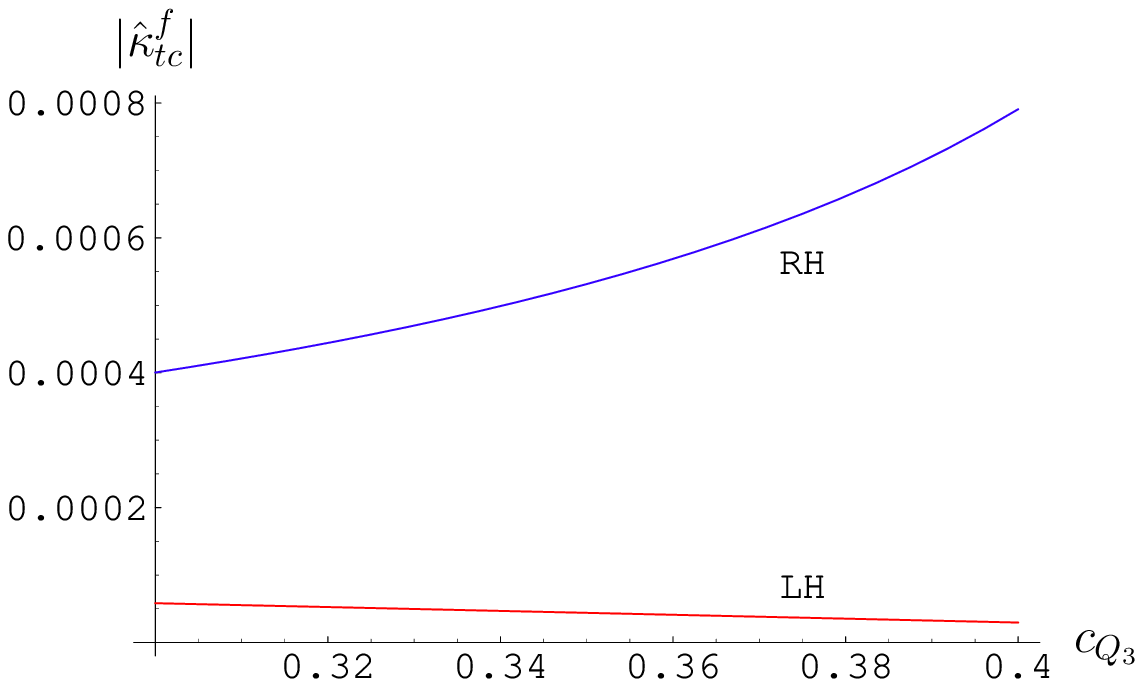}}
\caption{\label{Fig:hkftuc} Fermion KK contribution in the case of symmetrical
mass matrices to (a) $\hat{\kappa}_{tu}$ and (b) $\hat{\kappa}_{tc}$. The 
labels LH and RH indicate whether it is for the LH or RH coupling. The plots 
are made using the first KK mode to approximate the full KK tower.}
\end{figure}

For the fermion KK contributions, since the decoupling of the higher KK modes
is very efficient, hence just the first KK mode provides a very good
approximation to the full tower. We plot using this approximation
$|\hat{\kappa}^{f}_{3r}|$ as functions of $c_{Q_3}$ in Fig.~\ref{Fig:hkftuc}.

\subsection{\label{Sec:tcZ} Experimental signatures at the LHC}
\begin{figure}[htbp]
\centering
\subfigure[]{
\label{Fig:subfig:tuZ}
\includegraphics[width=2.8in]{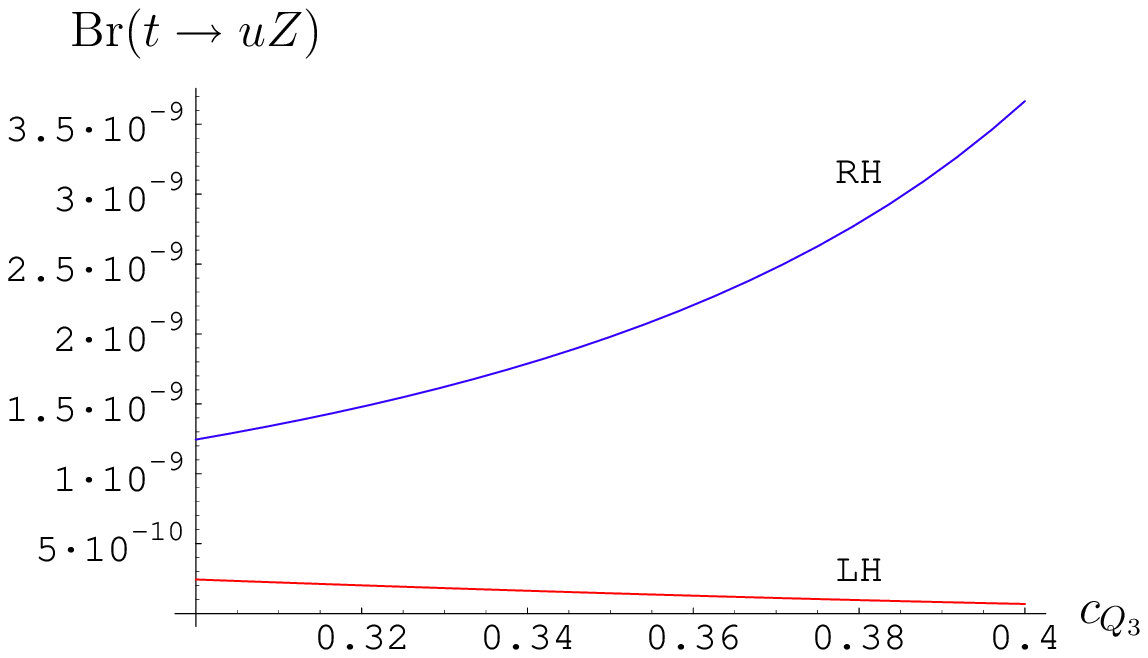}}
\hspace{0.2in}
\subfigure[]{
\label{Fig:subfig:tcZ}
\includegraphics[width=2.8in]{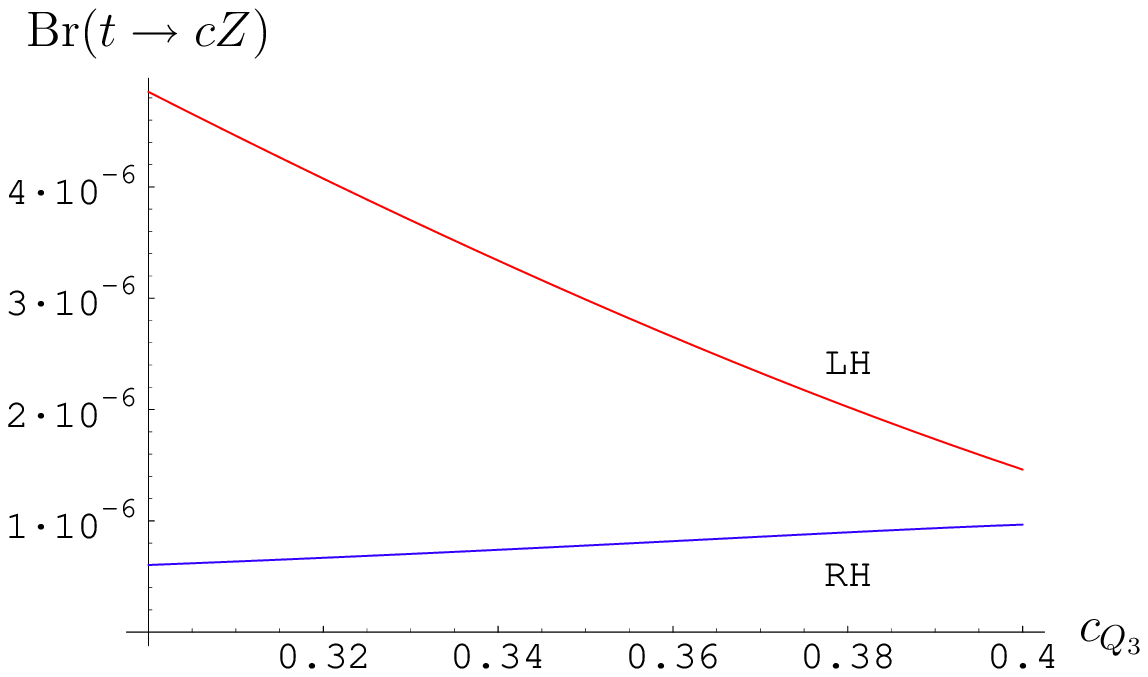}}
\caption{\label{Fig:BrsymM} Branching ratio in the case of symmetrical mass 
matrices as a function of $c_{Q_3}$ for the decay (a) $t \ra u\,Z$ and (b) 
$t \ra c\,Z$. The labels LH and RH indicate LH or RH top decay.}
\end{figure}

The branching ratio of the decay $t \ra c(u) Z$ is given by
\begin{align}\label{Eq:BrtcZ}
\mathrm{Br}(t \ra c(u) Z) &= \frac{2}{c^2}
\Big(|Q_Z(t_L)\,\hat{\kappa}^L_{tc(u)}|^2 +
|Q_Z(t_R)\,\hat{\kappa}^R_{tc(u)}|^2\Big)
\left(\frac{1-x_t}{1-y_t}\right)^2
\left(\frac{1+2x_t}{1+2y_t}\right)\frac{y_t}{x_t} \,, 
\end{align}
where $x_t= m_Z^2/m_t^2$ and $y_t=m_W^2/m_t^2$. In Fig.~\ref{Fig:BrsymM} we
plot the branching ratio as a function of $c_{Q_3}$ in the case where the weak
eigenbasis mass matrix has the symmetric ansatz form of~\eqref{Eq:MNM}.
It is clear that the dominant channel is $t \ra c\,Z$. The branching ratio
is at the level of a few $10^{-6}$, which is to be compared to the SM
prediction of $\mathcal{O}(10^{-13})$~\cite{SMt}. As $c_{Q_3}$ increases, the
decay changes from being mostly coming from the LH tops at the low end of the
allowed range of $c_{Q_3}$, to having comparable contributions from both quark
helicities at the high end. Note that one can in principle differentiate
whether the quark rotation is LH or RH by studying the polarized top decays.

For the case of asymmetrical quark mass matrix configurations found in
Sec.~\ref{Sec:Rand}, the resultant branching ratios and the associated gauge
and fermion KK flavour off-diagonal contributions are tabulated in
Table~\ref{Tb:ASBrtc}. We give results only for the decay into charm quarks
since this channel dominates over that into the up-quarks. The magnitude of
our branching ratios for both cases of symmetrical and asymmetrical quark mass
matrices are consistent with previous estimate in the RS
framework~\cite{APS07}.
\begin{table}[htbp]
\caption{\label{Tb:ASBrtc} Branching ratios of $t \ra c\,Z$ and the associated
gauge and fermion KK flavour off-diagonal contributions for the case of
asymmetrical mass matrices found from numerical searches.}
\begin{ruledtabular}
\begin{tabular}{ccccccc}
Config. & $|\hat{\kappa}^g_L|$ & $|\hat{\kappa}^g_R|$ & $|\hat{\kappa}^f_L|$ &
$|\hat{\kappa}^f_R|$ & Br($t_L$) & Br($t_R$) \\
\hline
I & $3.5 \times 10^{-4}$ & $7.7 \times 10^{-3}$ & $8.2 \times 10^{-3}$ &
$4.7 \times 10^{-3}$ & $1.4 \times 10^{-5}$ & $4.1 \times 10^{-6}$ \\
II & $4.3 \times10^{-4}$ & $5.8 \times 10^{-3}$ & $9.9 \times 10^{-3}$ &
$2.9 \times 10^{-3}$ & $2.1 \times 10^{-5}$ & $2.0 \times10^{-6}$ \\
III & $2.1 \times10^{-4}$ & $3.8 \times 10^{-3}$ & $5.0 \times 10^{-3}$ &
$7.0 \times 10^{-3}$ & $5.4 \times 10^{-6}$ & $3.2 \times 10^{-6}$ \\
\end{tabular}
\end{ruledtabular}
\end{table}

It is interesting to note from Fig.~\ref{Fig:BrsymM}(b) and 
Table~\ref{Tb:ASBrtc} that in $t \ra c\,Z$ decays, the LH decays dominate over
the RH ones in the case of both symmetrical and asymmetrical quark mass 
matrices. The reason for this is however different for the two cases. 
In the symmetric case, $M_u = M_u^\dag$ and so $U^u_L = U^u_R = U^u$. Thus the
difference between the LH and RH decays is due to the differences in the weak 
eigenbasis couplings, as can be seen from Eq.~\eqref{Eq:hatkg}, and $Q_Z$. By 
comparing Fig.~\ref{Fig:hkftuc}(b) to~\ref{Fig:hkgtuc}(b) we see 
$|\hat{\kappa}_{tc}|\sim|\hat{\kappa}^g_{tc}|$, and from 
Fig.~\ref{Fig:hkgtuc}(b) we have 
$0.9 \lesssim |(\hat{\kappa}^g_{tc})_R|/|(\hat{\kappa}^g_{tc})_L| \lesssim 2$~\footnote{It may seem counterintuitive that $|(\hat{\kappa}^g_{tc})_R|$ can be 
smaller than $|(\hat{\kappa}^g_{tc})_L|$ (for $c_{Q_3} < 0.32$), as one may 
expect that the couplings to be dominated by the top contribution, and the 
coupling to the RH top to be larger than that to the LH top due to the fact that
the RH top is localized closer to the IR brane. However, such expectations can
be misleading. Because of the mixing matrices, the mass eigenbasis coupling, 
$\hat{\kappa}^g_{tc}$, is not just a simple sum of the weak eigenbasis 
couplings, $\kappa^g_{q_i}$, but involves their differences as already 
mentioned. Moreover, although the greatest contribution comes from the top, 
the contribution from the second generation may not be completely negligible, 
as is the case here for $(\kappa^g_c)_R$ for the particular symmetric ansatz 
that we study.}.
However, as $|Q_Z(t_L)|\gtrsim|2Q_Z(t_R)|$, the net effect is that the LH decay
dominates (see Eq.~\eqref{Eq:BrtcZ}).

In the asymmetrical case, $M_u \neq M_u^\dag$ and $U^u_L \neq U^u_R$ with no 
pattern relating the LH to the RH mixings. In each of the configurations of 
localization parameters listed in Table~\ref{Tb:ASBrtc}, while  
$|(\hat{\kappa}^g_{tc})_L| \ll |(\hat{\kappa}^g_{tc})_R|$, it turns out that 
not only $|(\hat{\kappa}^g_{tc})_R|\sim|(\hat{\kappa}^f_{tc})_R|$ and
$|(\hat{\kappa}^f_{tc})_L|\sim|(\hat{\kappa}^f_{tc})_R|$, there is also a
relative minus sign between the gauge and the fermion KK contributions, which
results in a destructive interference that leads to a greater branching ratio
for the LH decay. This is to be contrasted with Ref.~\cite{APS07} where it is 
the RH mode that is found to dominate. There it appears that the possibility of
having a cancellation between the gauge and fermion KK contributions was not 
considered. 

We note and emphasize here the crucial role the quark mass and mixing matrices
play in determining the mass eigenbasis flavour off-diagonal couplings 
$\hat{\kappa}_{ab}$. Most importantly, $\hat{\kappa}_{ab}$ do not depend on 
the fermion localizations alone. Whether or not there is a cancellation between
the gauge and fermion KK contributions depends very much on the combination of
the particular quark mass and mixing matrices considered just as well as the 
configuration of fermion localizations used. Such cancellation is by no mean 
generic, and has to be checked whenever a new combination of admissible
 configuration of fermion localizations, and quark mass and mixing matrices 
arise. In addition, since 5D gauge and Yukawa couplings are independent 
parameters, whether or not 
$|(\hat{\kappa}^g_{ab})_L| \ll |(\hat{\kappa}^g_{ab})_R|$ does not mean the 
same has to hold between $|(\hat{\kappa}^f_{ab})_L|$ and 
$|(\hat{\kappa}^f_{ab})_R|$. Since $\kappa^g_{ij}$ and $\kappa^f_{ij}$ have
very different structures (see Eqs.~\eqref{Eq:dgig} and~\eqref{Eq:dgfg}), the 
combined effect when convolved with the particular quark mixing matrices can be
quite different, as is the case for the three asymmetrical configurations 
listed in Table~\ref{Tb:ASBrtc}.

It is expected that both the single top and the $\bar{t}t$ pair production
rates will be high at the LHC, with the latter about a factor of two higher
still than the former. To a small correction, the single tops are always 
produced in the LH helicity, while both helicities are produced in pair
productions. Thus a simple way of testing the above at the LHC is to compare
the decay rates of $t \ra Z$ + jets in single top production events (e.g. in
the associated $t\,W$ productions) to that from the pair productions, so that 
informations of both LH and RH decays can be extracted. Note that both the
single and pair production channels should give comparable branching ratios 
initially at the discovery stage. Of course, a higher branching ratio would be
obtained from pair productions after several years of measurements.

\section{\label{Sec:Conc} Summary}
We have performed a detailed study of the admissible forms of quark mass
matrices in the MCRS model which reproduce the experimentally well-determined
quark mass hierarchy and CKM mixing matrix, assuming a perturbative and
hierarchyless Yukawa structure that is not fine-tuned.

We arrived at the admissible forms in two different ways. In one we examined
several quark mass matrix ansatz which are constructed to fit the quark masses
and the CKM matrix. These ansatz have a high degree of symmetries built in
which allows the localization of the quarks (that give rise to the mass
hierarchy in the RS setting) to be analytically determined.
We found that the Koide-type symmetrical ansatz is compatible with the
assumption of a hierarchyless Yukawa structure in the MCRS model, but not the
Fritzsch-type hermitian ansatz. Because the ansatzed mass matrices are
symmetrical, both LH and RH quark mixing matrices are the same.

In the other way, no \textit{a priori} quark mass structures were assumed. A 
numerical multiparameter search for configurations of quark localization 
parameters and Yukawa couplings that give admissible quark mass matrices was 
performed. Admissible configurations were found after an extensive search. No 
discernible symmetries or pattern were found in the quark mass matrices for 
both the up-type and down-type quarks. The LH and RH mixing matrices are found
to be different as is expected given the asymmetrical form of the mass 
matrices.

We studied the possibility of differentiating between the case of symmetrical
and asymmetrical quark mass matrices from flavour changing top decays,
$t \ra Z$ + jets. We found the dominant mode of decay is that with a final
state charm jet. The total branching ratio is calculated to be $\sim 3$ to 
$5 \times 10^{-6}$ in the symmetrical case and $\sim 9 \times 10^{-6}$ to 
$2 \times 10^{-5}$ in the asymmetrical case. The signal is within reach of the
LHC which has been estimated to be $6.5\times 10^{-5}$ for a $5\sigma$ signal 
at $100\,\mathrm{fb}^{-1}$~\cite{Atlas}. However, the difference between the 
two cases may be difficult to discern.

We have also investigated the decay $t_R\ra b_R\,W$ as a large number of top
quarks are expected to be produced at the LHC. We found a branching ratio
at the level of $\mathcal{O}(10^{-5})$ is possible. Although the signal is not 
negligible, given the huge SM background, its detection is still a very 
challenging task, and a careful feasibility study is needed. This is beyond the
scope of the present paper.

\section{acknowledgements}
W.F.C. is grateful to the TRIUMF Theory group for their
hospitality when part of this work was completed. The research of
J.N.N. and J.M.S.W. is partially supported by the Natural Science
and Engineering Council of Canada. The work of W.F.C. is supported
by the Taiwan NSC under Grant No. 96-2112-M-007-020-MY3.

{\em Note added}: After the completion of this work, we became aware of 
Ref.~\cite{CFW08} which finds that flavour bounds from the $\Delta F = 2$
processes in the meson sector, in particular that from $\epsilon_K$, might
require the KK mass scale to be generically $\mathcal{O}(10)$~TeV in the MCRS 
model. We will show in an ensuing publication~\cite{future} that parameter 
space generically exists where KK mass scale of a few TeV is still consistent 
with all the flavour constraints from meson mixings, and that our conclusions 
with regard to the top decay in this work continue to hold.

\appendix

\section{\label{app:HermM}The Hermitian Mass Matrix Ansatz}
In this appendix we show that generically, the Fritzsch-type ansatz cannot be
accommodated in the RS framework without requiring a hierarchy in the 5D Yukawa
couplings. We consider below a general Hermitian mass matrix ansatz for which
the Fritzsch-type ansatz is a special case of.

\subsection{General analytical structure}
The Hermitian mass matrix ansatz takes the form
\begin{equation}\label{Eq:FXM}
M_f
= P_f^\hc \hat{M}_f P_f \,, \quad f = u,\,d,
\end{equation}
where $P_f = \mrm{diag}\{1,\,e^{i\phi_{C_f}},\,e^{i(\phi_{B_f}+\phi_{C_f})}\}$
is a diagonal pure phase matrix, and
\begin{equation}
\hat{M}_f =
\begin{pmatrix}
 U_f  & |C_f| &  V_f \\
|C_f| &  D_f  & |B_f| \\
 V_f^\star  & |B_f| &  A_f
\end{pmatrix} \,, \quad
V_f = |V_f|\,e^{i\omega_f} \,, \quad
\omega_f = \phi_{B_f}+\phi_{C_f}-\phi_{V_f} \,,
\end{equation}
with $\phi_X \equiv \mathrm{arg}(X)$ and
$A_f,\,D_f,\,U_f,\,|X|,\,\phi_X \in \mathbb{R}$. Note that the Fritzsch-type
ansatz with four texture zeroes~\cite{FX03} is recovered when $U_f = V_f = 0$
(the six-zero texture case~\cite{Fansatz} has $D_f = 0$ also). For simplicity,
we take $\omega_f\in\{0,\,\pi\}$ below so that
$V_f = \pm|V_f|$.~\footnote{Such case has been considered in Ref.~\cite{GM07},
and was shown to be consistent with the current experimental CKM data.}
We will ignore the fermion label below for convenience.

The matrix $\hat{M}$ can be diagonalized via an orthogonal transformation
\begin{equation}
O^\mrm{T} \hat{M} O =
\begin{pmatrix}
\lambda_1 & 0         & 0 \\
0         & \lambda_2 & 0 \\
0         & 0         & \lambda_3
\end{pmatrix} \,, \quad
|\lambda_1| < |\lambda_2| < |\lambda_3| \,.
\end{equation}
The eigenvalues $|\lambda_i|$, $i = 1,\,2,\,3$, can be either positive or
negative. To reproduce the observed mass spectrum, we set $|\lambda_i| = m_i$.
From the observed quark mass hierarchy, it is expected in general that $|A|$
be the largest entries in $\hat{M}$, and $|A|\lesssim|\lambda_3|$. Without
loss of generality, we take $A$ and $\lambda_3$ to be positive.

By applying the Cayley-Hamilton theorem, three independent relations between
the six parameters of $\hat{M}$ to its three eigenvalues can be deduced:
\begin{align}\label{Eq:BCDreln}
&S_1 - A - D - U = 0 \,, &
S_1 &= \sum_i\lambda_i \,, \notag \\
&S_2 + |B|^2 + |C|^2 + V^2 - A D - (A + D)U = 0 \,, &
S_2 &= \sum_{i<j}\lambda_i\lambda_j \,, \notag \\
&S_3 + A|C|^2 + D V^2 + U|B|^2 - A D U - 2|B||C|V = 0 \,, &
S_3 &= \prod_i\lambda_i \,.
\end{align}
Choosing $A$, $U$, $V$ to be the free parameters, Eq.~\eqref{Eq:BCDreln} can be
solved for $|B|$, $|C|$, and $D$:
\begin{gather}\label{Eq:BCDsoln}
D = S_1 - A - U \,, \notag \\
|B| = \frac{VY + Z}{\sqrt{(A-U)X - 2V(VY + Z)}} \,, \qquad
|C| = \sqrt{\frac{(A-U)X - 2V(VY - Z)}{(A-U)^2 + 4V^2}} \,,
\end{gather}
where
\begin{align}
X &= U^3 + (A + 2U)V^2 - (U^2+V^2)S_1 + U S_2 - S_3 \,, \notag \\
Y &= A^2 + V^2 + (A + U)(U-S_1) + S_2 \,, \notag \\
Z &= \sqrt{V^2 Y^2 + (U-A)X Y - X^2} \,.
\end{align}
If $|U|,\,|V| \ll |\lambda_1| \ll |A|$ so that Eq.~\eqref{Eq:FXM} is a
perturbation of the Fritzsch four-zero texture ansatz, $|B|$ and $|C|$ can be
expanded as
\begin{align}
|B| &= \pm\sqrt{\frac{-\prod_i(A-\lambda_i)}{A}}\left[
1+\frac{\eps_U}{2}\mp\eps_V R+\mathcal{O}(\eps_U^2,\,\eps_V^2)
\right] \notag \\
&\qquad +\frac{V A^{3/2}}{\sqrt{-S_3}}\left(
1-\frac{S_1}{A}+\frac{S_2}{A^2}\right)\left[
1+\frac{U S_2}{2S_3}+\frac{\eps_U}{2}\mp\eps_V R(A)
+\mathcal{O}(\eps_U^2,\,\eps_V^2)\right] \,, \notag \\
|C| &= \sqrt{\frac{-S_3}{A}}\left[
1-\frac{U S_2}{2S_3}+\frac{\eps_U}{2}
\mp\eps_V R+\mathcal{O}(\eps_U^2,\,\eps_V^2)
\right] \,,
\end{align}
where
\begin{equation}
\eps_U = \frac{U}{A} \,, \quad \eps_V = \frac{V}{A} \,, \quad
R = \sqrt{\frac{\prod_i(A-\lambda_i)}{S_3}} \,.
\end{equation}
Given that we have taken $A < \lambda_3$ and $A,\,\lambda_3 > 0$, it is
required that $S_3 < 0$ (or $\lambda_1\lambda_2 < 0$) for $|B|$ and $|C|$ to
be real. This is consistent with the expectation from the considerations of
Ref.~\cite{FX03}. In the limit $U,\,V \to 0$, the exact Fritzsch four-zero
texture ansatz is recovered.

With $\hat{M}$ determined by the three free parameters which we chose to be
$A$, $U$, and $V$, so are its eigenvectors. For each eigenvalue $\lambda_i$,
its associated eigenvector takes the form
\begin{equation}
\mathcal{\bsym{v}}_i^\mrm{T} = \Big(
|C|(A-\lambda_i)-|B|V \,,\,
V^2-(A-\lambda_i)(U-\lambda_i) \,,\,
|B|(U-\lambda_i)-|C|V
\Big)^\mrm{T} \,.
\end{equation}
The orthogonal matrix $O$ is then given by
\begin{equation}
O =
\begin{pmatrix}
| & | & | \\
\bar{\mathcal{\bsym{v}}}_1 & \bar{\mathcal{\bsym{v}}}_2
& \bar{\mathcal{\bsym{v}}}_3 \\
| & | & |
\end{pmatrix} \,, \qquad
\bar{\mathcal{\bsym{v}}}_i\equiv
\frac{\mathcal{\bsym{v}}_i}{\|\mathcal{\bsym{v}}_i\|} \,, \quad
i = 1,\,2,\,3 \,,
\end{equation}
and the quark mixing matrix by
$V_\mrm{mix} \equiv O_u^\mrm{T}(P_u P_d^\dag)O_d$.

\subsection{Matching to the RS mass matrix}
To reproduce the Hermitian mass matrix ansatz~\eqref{Eq:FXM} by the RS mass
matrix~\eqref{Eq:RSM}, we match them and solve for the parameters determining
the RS mass matrix. For the purpose of checking if hierarchy arises in the 5D
Yukawa couplings from the matching, we may start matching in either the up or
the down sector. For simplicity, the fermion species label is ignored below.

There are a total of 24 parameters in $M^{RS}$ to be determined: six fermion
wave function values, $F_L(c^L_i)$ and $F_R(c^R_i)$, nine Yukawa magnitudes,
$\rho_{ij}$, and nine Yukawa phases, $\phi_{ij}$, where $i,j = 1,\,2,\,3$.
Matching results in nine conditions for both the magnitudes and the phases. Thus
all the Yukawa phases are determined by $\phi_{B,C}$, while six magnitudes are
left as free independent parameters. These we chose to be $F_L(c^L_3)$ and
$F_R(c^R_3)$, which are constrained by EWPT, and $\rho_{11}$, $\rho_{21}$,
$\rho_{31}$, $\rho_{32}$. The determined parameters are then the five Yukawa
magnitudes:
\begin{gather}
\rho_{13} = \frac{kL}{F_L(c_3^L)F_R(c_3^R)}
\frac{V^2}{v_W U}\frac{\rho_{11}}{\rho_{31}} \,, \qquad
\rho_{23} = \frac{kL}{F_L(c_3^L)F_R(c_3^R)}
\frac{V\,|B|}{v_W|C|}\frac{\rho_{21}}{\rho_{31}} \,, \notag \\
\rho_{33} = \frac{kL}{F_L(c_3^L)F_R(c_3^R)}\frac{A}{v_W} \,, \notag \\
\rho_{12} = \frac{|C|\,V}{|B|\,U}
\frac{\rho_{11}\,\rho_{32}}{\rho_{31}} \,, \qquad
\rho_{22} = \frac{D\,V}{|B||C|}\frac{\rho_{21}\,\rho_{32}}{\rho_{31}} \,,
\label{Eq:FXrho}
\end{gather}
the nine Yukawa phases:
\begin{align}
\phi_{11} &= 0 \,, &
\phi_{12} &= \phi_C \,, &
\phi_{13} &= \phi_B+\phi_C \,, \notag \\
\phi_{21} &= -\phi_C \,, &
\phi_{22} &= 0 \,, &
\phi_{23} &= \phi_B \,,\notag \\
\phi_{31} &= -\phi_B-\phi_C \,, &
\phi_{32} &= -\phi_B \,, &
\phi_{33} &= 0 \,,
\end{align}
and the four fermion wave function values:
\begin{align}
F_L(c^L_1) &= F_L(c^L_3)\frac{U}{V}\frac{\rho_{31}}{\rho_{11}} \,, &
F_L(c^L_2) &= F_L(c^L_3)\frac{|C|}{V}\frac{\rho_{31}}{\rho_{21}} \,,
\notag \\
F_R(c^R_1) &= \frac{V}{v_W}\frac{kL}{F_L(c^L_3)\rho_{31}} \,, &
F_R(c^R_2) &= \frac{|B|}{v_W}\frac{kL}{F_L(c^L_3)\rho_{32}} \,.
\end{align}
Note that there are only three independent Yukawa phases because the mass
matrix ansatz is Hermitian. Note also that since fermion wave functions are
always positive, $V$ and thus $U$ have to be positive implying that
$\omega = 0$.

From Eq.~\eqref{Eq:FXrho}, in order for the Yukawa couplings to be of the same
order, it is required that
$\rho_{11}\sim\rho_{21}\sim\rho_{31}\sim\rho_{32}\sim\rho_{33}$, and
\begin{gather}
\frac{|C|\,V}{|B|\,U} \sim 1 \,, \quad
\frac{D\,V}{|B||C|} \sim 1 \,, \quad
\frac{V^2}{U A} \sim 1 \quad \Longrightarrow \quad
\frac{V}{U}\sim\frac{A}{V}\sim\frac{|B|}{|C|} \,.
\end{gather}
For generic sets of parameters we find $|B_u|/|C_u|\sim\mathcal{O}(10^3)$
and $|B_d|/|C_d|\sim\mathcal{O}(50)$. However, parameter sets that reproduce
all entries of the CKM matrix and also the Jarlskog invariant to within two
standard error can only be found if $V_u \sim U_u$ and $V_d \sim 10U_d$. Thus
hierarchy in the 5D Yukawa couplings cannot be avoided if the Hermitian mass
matrix ansatz~\eqref{Eq:FXM} is to be accommodated in the RS framework.

\section{\label{app:SymmM}RS matching of the symmetric ansatz}
In this appendix, we give analytical expressions for the parameters determined
from matching the RS mass matrix~\eqref{Eq:RSM} to the mass matrix
ansatz~\eqref{Eq:MNM}.
Starting with the up sector, the determined parameters are the five up-type
Yukawa magnitudes:
\begin{gather}
\rho^u_{13} = \frac{kL}{F_L(c_{Q_3})F_R(c_{U_3})}\frac{C_u^2}{v_W\xi_u}
\frac{\rho^u_{11}}{\rho^u_{31}} \,,
\qquad
\rho^u_{23} = \frac{kL}{F_L(c_{Q_3})F_R(c_{U_3})}\frac{B_u}{v_W}
\frac{\rho^u_{21}}{\rho^u_{31}} \,,
\notag \\
\rho^u_{33} = \frac{kL}{F_L(c_{Q_3})F_R(c_{U_3})}\frac{A_u}{v_W} \,,
\notag \\
\rho^u_{12} = \frac{C_u^2}{B_u\xi_u}
\frac{\rho^u_{11}\rho^u_{32}}{\rho^u_{31}} \,,
\qquad
\rho^u_{22} = \frac{A_u}{B_u}\frac{\rho^u_{21}\rho^u_{32}}{\rho^u_{31}} \,,
\label{Eq:yQu}
\end{gather}
the nine up-type Yukawa phases:
\begin{align}\label{Eq:phQu}
\phi^u_{11} &= -2\delta^u_1 \,, &
\phi^u_{12} &= -\delta^u_1-\delta^u_2 \,, &
\phi^u_{13} &= -\delta^u_1-\delta^u_3 \,, \notag \\
\phi^u_{21} &= -\delta^u_1-\delta^u_2 \,, &
\phi^u_{22} &= -2\delta^u_2 \,, &
\phi^u_{23} &= -\delta^u_2-\delta^u_3 \,, \notag \\
\phi^u_{31} &= -\delta^u_1-\delta^u_3 \,, &
\phi^u_{32} &= -\delta^u_2-\delta^u_3 \,, &
\phi^u_{33} &= -2\delta^u_3 \,,
\end{align}
and the four fermion wave function values:
\begin{align}\label{Eq:FLQFRU}
F_L(c_{Q_1}) &= F_L(c_{Q_3})\frac{\xi_u}{C_u}\frac{\rho^u_{31}}{\rho^u_{11}}
\,, &
F_L(c_{Q_2}) &= F_L(c_{Q_3})\frac{\rho^u_{31}}{\rho^u_{21}}
\,, \notag \\
F_R(c_{U_1}) &= \frac{kL}{F_L(c_{Q_3})}\frac{C_u}{v_W}\frac{1}{\rho^u_{31}}
\,, &
F_R(c_{U_2}) &= \frac{kL}{F_L(c_{Q_3})}\frac{B_u}{v_W}\frac{1}{\rho^u_{32}} \,.
\end{align}

Next the down sector. Given the information on the up sector, the determined
parameters are the six down-type Yukawa magnitudes:
\begin{align}\label{Eq:yQd}
\rho^d_{11} &= \frac{F_L(c_{Q_3})}{F_L(c_{Q_1})}\frac{\xi_d}{C_d}\rho^d_{31}
= \frac{C_u}{\xi_u}\frac{\xi_d}{C_d}\frac{\rho^u_{11}}{\rho^u_{31}}\rho^d_{31}
\,, &
\rho^d_{21} &= \frac{F_L(c_{Q_3})}{F_L(c_{Q_2})}\rho^d_{32}
= \frac{\rho^u_{21}}{\rho^u_{31}}\rho^d_{31}
\,, \notag \\
\rho^d_{12} &= \frac{F_L(c_{Q_3})}{F_L(c_{Q_1})}\frac{C_d}{|B_d|}\rho^d_{32}
= \frac{C_u}{\xi_u}\frac{C_d}{|B_d|}\frac{\rho^u_{11}}{\rho^u_{31}}\rho^d_{32}
\,, &
\rho^d_{22} &= \frac{F_L(c_{Q_3})}{F_L(c_{Q_2})}\frac{A_d}{|B_d|}\rho^d_{32}
= \frac{A_d}{|B_d|}\frac{\rho^u_{21}}{\rho^u_{31}}\rho^d_{32}
\,, \notag \\
\rho^d_{13} &= \frac{F_L(c_{Q_3})}{F_L(c_{Q_1})}\frac{C_d}{A_d}\rho^d_{33}
= \frac{C_u}{\xi_u}\frac{C_d}{A_d}\frac{\rho^u_{11}}{\rho^u_{31}}\rho^d_{33}
\,, &
\rho^d_{23} &= \frac{F_L(c_{Q_3})}{F_L(c_{Q_2})}\frac{|B_d|}{A_d}\rho^d_{33}
= \frac{|B_d|}{A_d}\frac{\rho^u_{21}}{\rho^u_{31}}\rho^d_{33} \,,
\end{align}
the nine down-type Yukawa phases:
\begin{align}\label{Eq:phQd}
\phi^d_{11} &= -2\delta^d_1 \,, &
\phi^d_{12} &= -\delta^d_1-\delta^d_2 \,, &
\phi^d_{13} &= -\delta^d_1-\delta^d_3 \,, \notag \\
\phi^d_{21} &= -\delta^d_1-\delta^d_2 \,, &
\phi^d_{22} &= -2\delta^d_2 \,, &
\phi^d_{23} &= \pi-\delta^d_2-\delta^d_3 \,, \notag \\
\phi^d_{31} &= -\delta^d_1-\delta^d_3 \,, &
\phi^d_{32} &= \pi-\delta^d_2-\delta^d_3 \,, &
\phi^d_{33} &= -2\delta^d_3 \,,
\end{align}
and the three fermion wave function values:
\begin{equation}\label{Eq:FRD}
F_R(c_{D_1}) = \frac{C_d}{v_W}\frac{kL}{F_L(c_{Q_3})}\frac{1}{\rho^d_{31}}
\,, \quad
F_R(c_{D_2}) = \frac{|B_d|}{v_W}\frac{kL}{F_L(c_{Q_3})}\frac{1}{\rho^d_{32}}
\,, \quad
F_R(c_{D_3}) = \frac{A_d}{v_W}\frac{kL}{F_L(c_{Q_3})}\frac{1}{\rho^d_{33}}
\,.
\end{equation}

Note that there are only six independent up-type Yukawa phases and six for the
down-type Yukawa phases since the mass matrix ansatz is symmetric. With the
texture phases $\delta^f_{1,2,3}$ determined by fitting the CKM data, there
are three more relations, i.e. $\delta^d_1 = \delta^u_1 = 0$ and
$\delta_{2,3} = \delta^u_{2,3}-\delta^d_{2,3}$, which further reduce the
number of independent Yukawa phases from a total of 12 down to nine.

In order to be consistent with EWPT
($\delta g_{Z b_L\bar{b}_L}/g_{Z b_L\bar{b}_L} \lesssim 0.01$~\footnote{The 
bound we adopted here is that from the PDG. Studies of similar model but 
differing details where a complete electroweak analysis was carried out have 
produced a more stringent bound, e.g. $\lesssim 0.0025$~\cite{AC06}. Such 
complete EWPT analysis, however, is beyond the scope of the present work.}) 
and to avoid too large a correction to the Peskin-Takeuchi S and T parameters,
it is required that $0.25 < c_{Q_3} < 0.4$, $c_{U_3} < 0.2$, so that 
$m^{(1)}_{gauge} \lesssim 4$~TeV~\cite{ADMS03}. To have the theory weakly 
coupled for at least the first two KK modes, $|\lambda_5| < 4$ is required 
also~\cite{APS05}. It follows that $2.70 < F_L(c_{Q_3}) < 4.27$,
$F_R(c_{U_3}) < 7.15$, and $\rho^{u,d}_{ij} < 4$, which when combined with
Eqs.~\eqref{Eq:yQu}, \eqref{Eq:yQd}, \eqref{Eq:CKMfit} and~\eqref{Eq:ABCnum}
imply
\begin{gather}\label{Eq:constr1}
4.06 < F_L(c_{Q_3})F_R(c_{U_3}) < 30.57 \,, \qquad
0.53 < \rho^u_{33} < 4 \,,
\end{gather}
and
\begin{equation}\label{Eq:constr2}
\frac{\rho^u_{11}}{\rho^u_{31}} < 0.14F_L(c_{Q_3})F_R(c_{U_3}) \,, \qquad
\frac{\rho^u_{21}}{\rho^u_{31}} < 0.25F_L(c_{Q_3})F_R(c_{U_3}) \,,
\end{equation}
\begin{align}\label{Eq:constr3}
\frac{\rho^u_{11}}{\rho^u_{31}}\rho^u_{32} &< 2.19 \,, &
\frac{\rho^u_{11}}{\rho^u_{31}}\rho^d_{31} &< 1.52 \,, &
\frac{\rho^u_{11}}{\rho^u_{31}}\rho^d_{32} &< 1.92 \,, &
\frac{\rho^u_{11}}{\rho^u_{31}}\rho^d_{33} &< 1.99 \,, \notag \\
\frac{\rho^u_{21}}{\rho^u_{31}}\rho^u_{32} &< 3.97 \,, &
\frac{\rho^u_{21}}{\rho^u_{31}}\rho^d_{31} &< 4    \,, &
\frac{\rho^u_{21}}{\rho^u_{31}}\rho^d_{32} &< 3.86 \,, &
\frac{\rho^u_{21}}{\rho^u_{31}}\rho^d_{33} &< 4.15 \,.
\end{align}

\begin{figure}[htbp]
\centering
\includegraphics[width=3.5in]{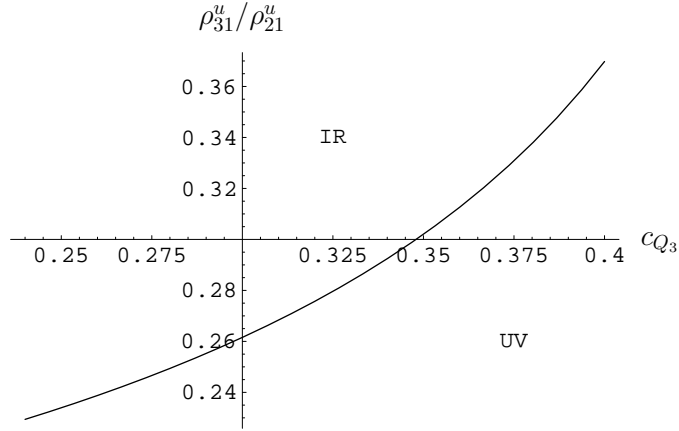}
\caption{\label{Fig:cQ2ru} The critical value of $\rho^u_{31}/\rho^u_{21}$ as
a function of $c_{Q_3}$ in the range allowed by EWPT. For values of
$\rho^u_{31}/\rho^u_{21}$ in the ``UV'' (``IR'') region, $c_{Q_2}$ is greater
(less) than 0.5.}
\end{figure}

Observe from Eq.~\eqref{Eq:FLQFRU} that the second generation $SU(2)_L$
doublet, $Q_2$, is localized towards the UV (IR) brane if
$\rho^u_{31}/\rho^u_{21}$ is less (greater) than $F_L(0.5+\eps)/F_L(c_{Q_3})$
($F_L(0.5-\eps)/F_L(c_{Q_3})$). Note that
$F_L(0.5\pm\eps) \approx 1\mp\eps k r_c\pi$ for $\eps \ll 1/(2k r_c\pi)$. We
plot in Fig.~\ref{Fig:cQ2ru} the critical value of $\rho^u_{31}/\rho^u_{21}$
below (above) which $Q_2$ is localized towards UV (IR) brane. The same logic
shows that the first generation $SU(2)_L$ doublet, $Q_1$, is generically
localized towards the UV brane because of the suppression factor
$\xi_u/C_u \sim 10^{-2}$ (even if $\rho^u_{31}/\rho^u_{11} \gtrsim 1$).

\end{document}